\documentclass[aps,pra,reprint,superscriptaddress,amsfonts,amssymb,amsmath]{revtex4-1}
\usepackage{graphicx,bm}
\usepackage[suffix=]{epstopdf}

\usepackage[caption=false]{subfig}
\newcommand{\eref}[1]{equation~\ref{#1}}
\newcommand{\sref}[1]{section~\ref{#1}}
\newcommand{\fref}[1]{figure~\ref{#1}}

\newcommand{\Sref}[1]{Section~\ref{#1}}
\newcommand{\Fref}[1]{Figure~\ref{#1}}

\begin{document}
\title{Physical realization of complex dynamical pattern formation in magnetic active feedback rings}

\author{Justin Q. Anderson}
\email{jusander@mines.edu}
\affiliation{Department of Physics, Colorado School of Mines, Golden, Colorado 80401, USA}
\author{P. A. Praveen Janantha}
\affiliation{Department of Physics, Colorado State University, Fort Collins, Colorado 80523, USA}
\author{Diego A. Alcala}
\affiliation{Department of Physics, Colorado School of Mines, Golden, Colorado 80401, USA}
\author{Mingzhong Wu}
\affiliation{Department of Physics, Colorado State University, Fort Collins, Colorado 80523, USA}
\author{Lincoln D. Carr}
\affiliation{Department of Physics, Colorado School of Mines, Golden, Colorado 80401, USA}

\date{\today}

\begin{abstract}
We report the clean experimental realization of cubic-quintic complex Ginzburg-Landau physics in a single driven, damped system. Four numerically predicted categories of complex dynamical behavior and pattern formation are identified for bright and dark solitary waves propagating around an active magnetic thin film-based feedback ring: (1) periodic breathing; (2) complex recurrence; (3) spontaneous spatial shifting; and (4) intermittency. These nontransient, long lifetime behaviors are observed in microwave spin wave envelopes circulating within a dispersive, nonlinear yttrium iron garnet waveguide operating in a ring geometry where the net losses are directly compensated for via linear amplification on each round trip $\mathcal{O}(100~\mathrm{ns})$. The behaviors exhibit periods ranging from tens to thousands of round trip times $\mathcal{O}(\mathrm{\mu s})$ and are stable for 1000s of periods $\mathcal{O}(\mathrm{ms})$. We present 10 observations of these dynamical behaviors which span the experimentally accessible ranges of attractive cubic nonlinearity, dispersion, and external field strength that support the self-generation of backward volume spin waves in a four-wave-mixing dominant regime. Three-wave splitting is not explicitly forbidden and is treated as an additional source of nonlinear losses. These long lifetime behaviors of bright solitary waves span the categories of dynamical behavior previously numerically predicted to be observable and represent a complete experimental verification of the cubic-quintic complex Ginzburg-Landau equation as a model for the study of fundamental, complex nonlinear dynamics for driven, damped waves evolving in nonlinear, dispersive systems. These observed behaviors are persistent over long times and robust over wide parameter regimes, making them very promising for technological applications. The dynamical pattern formation of self-generated dark solitary waves in attractive nonlinearity, however, is entirely novel and is reported for both the periodic breather and complex recurrence behaviors. All behaviors are identified in the group velocity co-moving frame. For (1) periodic breathing, we find that four or fewer bright or dark solitary waves may exhibit breathing with stable periods ranging from tens to hundreds of round trip times. The location of the solitary waves within the ring are seen to shift predictably while maintaining both peak solitary wave amplitudes and widths. For (2) complex recurrence, we find the periodic recurrence of interactions of three or more bright or dark solitary wave peaks is observed with stable recurrence times varying from hundreds to tens of thousands of round trips. For (3) spontaneous spatial shifting, we find spontaneous relocation of otherwise stable underlying ring dynamics is characterized by the instantaneous shift in location, in the group velocity frame, of the solitary waves while maintaining all other characteristics of the behavior. The time between shifts is unpredictable. Finally, for (4) intermittency, the dynamical behavior observed within the feedback ring shifts between two or more stable underlying behaviors unpredictably, indicating the presence of two or more overlapping attractors within the system.
\end{abstract}

\maketitle

\section{\label{sec:intro}Introduction and Motivation}

Spin wave envelope (SWE) solitary waves in active magnetic thin film-based feedback rings (AFRs) have proven to be an effective sandbox for the exploration of fundamental nonlinear dynamics. Over the past two decades a rich variety of complex dynamical behaviors have been observed in dissipative SWE solitary waves propagating in these nonlinear, dispersive feedback rings. Examples include bright and dark solitary waves and wave trains~\cite{scott2005,Wu2004,Kalinikos1999,Slavin1994}, m\"{o}bius solitons~\cite{Demokritov2003}, Fermi-Pasta-Ulam and spatial recurrences~\cite{Wu2007,Scott2003}, chaotic solitary waves~\cite{Wang2011,Ustinov2011}, and random solitons~\cite{Wu2006,Tong2010}.

AFRs are a notably useful system for the study of dissipative, nonlinear, dispersive wave dynamics for a few fundamental reasons. First, the active feedback allows for the compensation, on average, of the losses that the SWE solitary waves experience during propagation. Such quasi-conservative evolution allows for the observation of dynamical behaviors which can persist for tens of thousands, or more, of the fundamental round trip time  $\mathcal{O}(100~\mathrm{ns})$. The experimental realization of such long life time, $\mathcal{O}(\mathrm{ms})$, dynamics is in fact in line with the original goals of solitary wave research which focused on the viability of solitons as the basis for long-distance high-bandwidth optical communication~\cite{Haus1996,Hasegawa2000,Ablowitz2000}. Research efforts in this area slowed in the mid 2000s as dispersion-managed solitary waves were ultimately abandoned in favor of multiplexing schemes which provided cheaper scalability. The varied ecology of long lifetime dynamical behaviors present in these systems and the accessibility of chaotic regimes, however, continues to make solitary waves an intriguing candidate for lower bandwidth secure communication~\cite{Amiri2015}.

Second, the feedback ring geometry imposes a phase constraint in the form of a ring resonance. Geometries of this type are common within the optics, electromagnetic device and magnonics communities~\cite{Nikitov2015,Chumak2015,Ustinov2018,Yi2018}. Active research topics include optical solitary waves produced in micro-ring resonators~\cite{Marin-Palomo2017}. These solitary waves are used as a source of broadband (octave spanning) frequency combs for coherent parallel communication. The dynamics of dissipative optical solitary waves within micro-ring resonators is also being investigated~\cite{Kippenberg2018,Panajotov2014,Zhou2019,Lucas2017}.  Spin wave AFRs are additionally studied as delay lines for reservoir computing~\cite{Riou2019,Watt2019,Nakane2018}.

Third, the dispersion and nonlinearity characteristics of SWEs in AFRs are highly tunable via two easily accessible experimental parameters: external field orientation and amplitude~\cite{Stancil2009,Wu2010,Kalinikos1986}. By adjusting these parameters one can vary the nonlinearity from three-wave mixing to four-wave (Kerr) mixing and from attractive to repulsive. The sign and amplitude of dispersion may also be manipulated this way. This allows for the study of both dark and bright solitary waves dynamics in a single system.

Fourth, SWEs in AFRs are described by the cubic-quintic complex Ginzburg-Landau equation (CQCGL), which in its nondimensionalized form, see~\eref{eqn:model:GLNLS}, is a generalized nonlinear Schr\"{o}dinger equation~\cite{Kalinikos1986}. Various forms of the nonlinear Schr\"{o}dinger equation appear as governing models across a wide variety of physical systems. This includes the Lugiato-Lefever equation which describes lasers in nonlinear cavities~\cite{Lugiato1987}, the Gross-Pitaevskii equation for modeling the mean field of atomic and molecular Bose-Einstein condensates~\cite{Pitaevskii2003,Carr2009}, and the Ginzburg-Landau equation which is used to describe superconductivity and the evolution of mode-locked laser envelopes~\cite{Arnason2002,Akhmediev2008}. Driven, damped nonlinear Schr\"{o}dinger equations are additionally used to model deep water waves~\cite{Sulem2004} as well as magnon and exciton-polariton Bose-Einstein condensation~\cite{Carusotto2013}.  The physics which are accessible using SWEs in AFRs are then, up to the nontrivial issues of units, scaling, complex potential and naming conventions, applicable to many physical systems which are described by this important family of isomorphic partial differential equations.  AFRs in fact provide an accessible tabletop experiment from which one can study complex nonlinear wave physics across many scales for driven, damped systems.

In this article we present the clean experimental realization of nontransient, long lifetime (10,000s of round trips) complex dynamical behaviors for SWE bright and dark solitary waves propagating within in an AFR. These results are distinct from the study of dissipative solitons in the above mentioned systems where focus has generally remained on exploring transient behaviors, periodic modulations and isolating extreme events~\cite{Chowdury2017,Yue2020,Serkin2019,Kevrekidis2019}. The behaviors described here were previously predicted to be observable by a numerical parameter space search which identified four distinct long lifetime examples of dynamical pattern formation in bright solitary waves described by the CQCGL~\cite{anderson2014complex}. We emphasize that these numerical predictions involved an extraordinarily broad parameter space search which spanned a minimum of \emph{five orders of magnitude} for four distinct parameters ($S, L , C$ and $Q$ in ~\eref{eqn:model:GLNLS}). Across that parameter space only four categories of long lifetime dynamical pattern formation were predicted. We report on the observation of all four of these behaviors for bright solitary waves and the first known realization of self-generation and dynamical pattern formation for dark solitary waves evolving under attractive nonlinearity.

These behaviors are promising for potential technological applications due to their persistently long lifetimes and robustness over wide parameter regimes. One likely application is classical benchmarking for quantum simulator experiments, for example Sagnac interferometers in BECs~\cite{helm2015sagnac}. Such devices have been explored in both attractive and repulsive nonlinearity regimes, but a lack of understanding of the attractive case~\cite{nguyen2017formation,oleks2019quantum,wales2020splitting} has prevented the field from moving forward technologically.

The observed dynamical behaviors are as follows. (1) We observe \emph{periodic breathing}, where solitary waves periodically relocate their position within the ring (in the group velocity frame). We present examples of one and two waves undergoing this periodic relocation for both bright and dark solitary waves. (2) We observe \emph{complex recurrence}, where three or more solitary waves undergo periodic spatial recurrence involving multiple frequencies. Examples of complex recurrence are shown for both bright and dark solitary waves. (3) We observe \emph{spontaneous spatial shifting}, where a stable behavior undergoes a sudden repositioning within the ring (in the group velocity frame) and then immediately continues its stable behavior. (4) Finally, we observe \emph{intermittency}, where the system jumps between two distinct stable behaviors.

The paper is organized as follows.~\Sref{sec:eandm} discusses the active magnetic thin film-based feedback ring experiment; here the experimental apparatus and accessible physics will be detailed. Sections \ref{sec:bre}-\ref{sec:int} focus on the key four behaviors as identified above: periodic breathers, complex recurrence, spontaneous spatial shifting and intermittency. Conclusions and outlooks are then provided in~\Sref{sec:con}.

\section{\label{sec:eandm}Experiment and Methods}

This section details the construction of the active magnetic thin film-based feedback ring used to collect the data reported in this work as well as the basic physics of the spin waves which we can access with it.  We start with a description of the AFR, its components and an overview of the merits and motivations behind the design of the experiment.  We then explore the relevant spin wave physics to the experiment and the subsequent analysis presented in this work.  Finally we discuss the experimental methods, procedure, and philosophy used to isolate the dynamics reported here.

\subsection{\label{sec:eandm:AFR}Active Magnetic thin film-based feedback Rings}

\begin{figure}
\includegraphics[width=.5\textwidth]{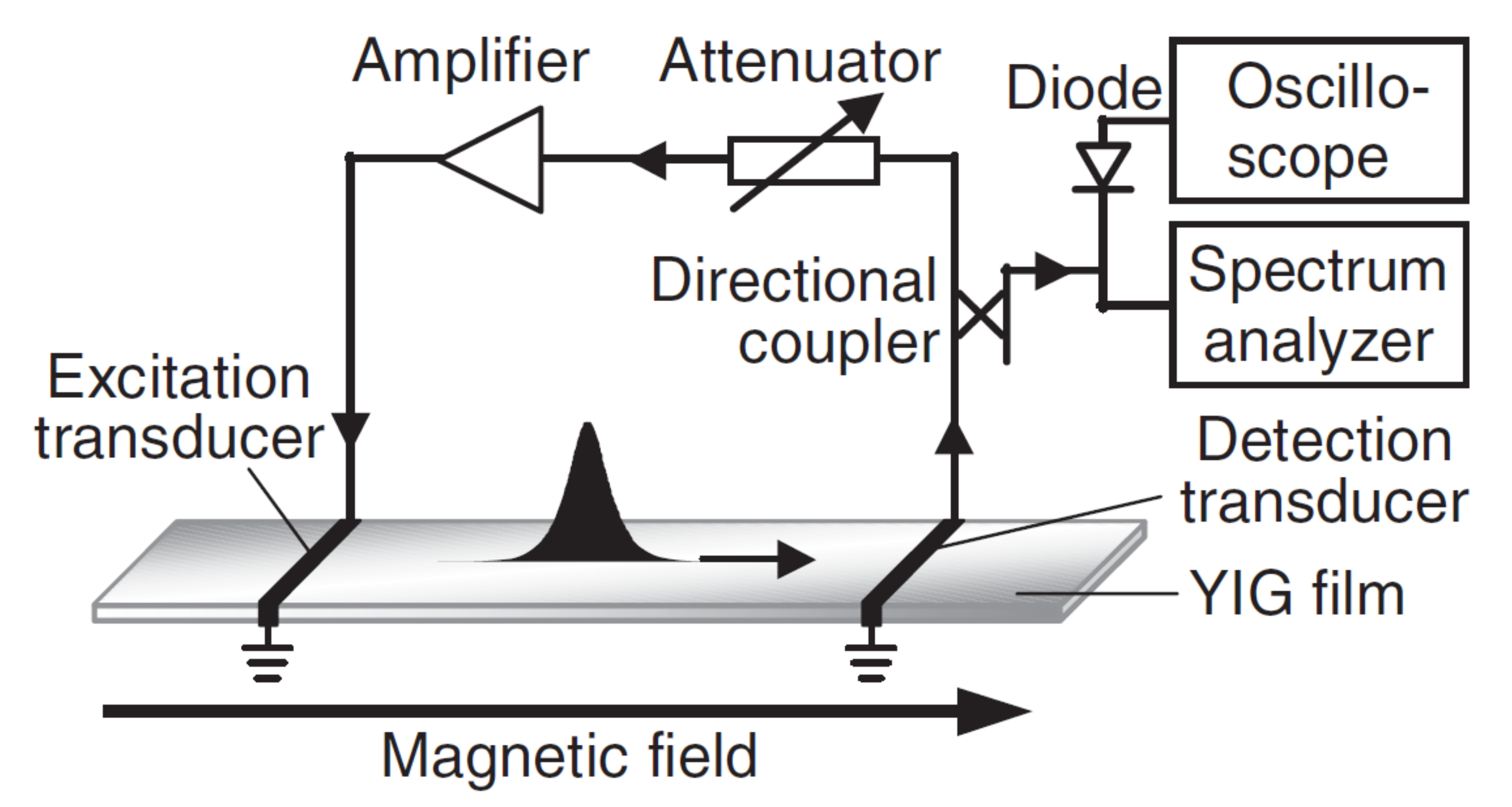}
\caption{\label{fig:exp_setup}\textit{Diagram of an active magnetic thin film-based feedback ring.} A spin wave feedback ring constructed of a nonlinear, dispersive waveguide (the YIG film) coupled to an electronic feedback loop via two transducers. The ring has a linear, fixed amplifier and a variable attenuator. Observation equipment is attached via a directional coupler. Reprinted with permission~\cite{Wu2006}.}
\end{figure}

An active magnetic thin film-based feedback ring is constructed of two main components: (1) a yttrium iron garnet (Y3Fe5O12,YIG) thin film waveguide and (2) an electronic feedback loop. These components are coupled via two transducers, one each for the excitation and detection of spin waves within the YIG waveguide. The electronic loop is comprised of a fixed, linear amplifier and variable attenuator pair which allows for the direct compensation of the major loss mechanisms present within the thin film. An oscilloscope and spectrum analyzer are coupled to the electronic feedback loop via a directional coupler for in-situ observation and recording of time-domain and frequency-domain ring signals. An AFR is schematically shown in~\fref{fig:exp_setup} detailing the construction of the electronic feedback loop and the thin film waveguide.~\Fref{fig:exp_setup} shows an external magnetic field applied parallel to the waveguide. This orientation enables the generation of backward volume spin waves (BVSW), the type of spin wave which is the main focus of this work. BVSWs will be discussed in detail in~\sref{sec:eandm:SW}, and any further mention of spin waves in this work will refer to BVSWs unless explicitly stated otherwise.

As mentioned in~\sref{sec:intro} the active feedback component of AFRs allows for the study of spin waves in a quasi-conservative regime, which can support behaviors with lifetimes of tens or hundreds of thousands of the fundamental round trip time, $\mathcal{O}(100~\mathrm{ns})$. The periodic amplification also enables the self-generation of SWE solitary waves. Self-generation is achieved by increasing the ring gain (through a decrease in attenuation) until the lowest loss ring eigenmode begins to circulate. The ring eigenmodes are given by the phase condition
\begin{equation}
k_n(\omega) L+\phi_e = 2\pi n,~~~n=1,2,3...,
\label{eqn:eignmodes}
\end{equation}
where $k_n(\omega)$ is the wavenumber of the $n^{\mathrm{th}}$ eigenmode, $L$ is the separation between the excitation and detection transducers and $\phi_e$ is a phase delay caused by the signal's propagation through the electronic feedback loop. Typically $\phi_e\ll k_n(\omega) L$. As the ring gain is increased additional eigenmodes begin to circulate and their amplitudes increase until these modes begin to interact nonlinearly through three and four-wave mixing processes (to be discussed in~\sref{sec:eandm:SW}). The number of eigenmodes present during a typical experiment is less than 10, though at higher ring gains up to a few hundred may circulate. All the solitary waves presented in this work are self-generated. Adjustments to the attenuator are done manually and occur on the order of seconds. This means all changes in ring gain occur over tens or hundreds of thousands of round trips and are adiabatic on that scale. Since no additional microwave signals are introduced into the ring from an outside source, all behaviors recorded from self-generated SWE solitary waves in AFRs have existed for at least~$\mathcal{O}(10^5)$ round trips and may be regarded as being in dynamic equilibrium. Transient behaviors are therefore not studied during self-generation. We note that transient effects can be highly relevant to applications such as communication, but they are not considered in this work.

Several other design features of the AFR warrant further discussions. First, YIG thin films are used as a propagation medium owing to their extraordinarily low loss characteristics at microwave frequencies~\cite{Wu2010}. These low losses, typically 3 orders lower than alternative waveguide materials, support propagation distances on the order of centimeters and therefore enable long excitation lifetimes. Long lifetimes maximize the interactions between nonlinearity, dispersion and losses on ring dynamics. Amplitude dependent nonlinearities and losses will have maximum relative influence on wave evolution close to the excitation transducer where spin wave power is at its highest. Dispersion and lower order effects will have higher relative influences the longer the spin wave propagates through the medium. 

Second, an external magnetic field is applied to magnetize the YIG thin film to saturation and thereby enable the excitation of spin waves. The direction and amplitude of this field allows for the tuning of nonlinearity and dispersion and will be further discussed in~\sref{sec:eandm:SW}. 

Third, all observation equipment is attached to the feedback ring via a directional coupler. This eliminates the interference of observation on ring dynamics. Observation may instead be treated as a fixed, linear attenuation within the AFR. 

Fourth, the signal is fed into the oscilloscope through a microwave diode with quadratic behavior. Thus, all recorded voltage data are taken to be proportional to power. The diode further allows for sampling of ring voltage at lower frequencies. This is because data recorded through the diode corresponds directly to the spin wave envelope, our principle interest, rather than the underlying carrier wave. These slower sampling rates of $\mathcal{O}(1~\mathrm{Gsamp/s})$ as opposed to~$\mathcal{O}(10~\mathrm{Gsamp/s})$ support the collection of longer time series which is crucial to the identification of dynamical behaviors which occur over hundreds or thousands of round trips. We do note that the reconstruction of phase is not possible without fully resolving the underlying carrier wave. Both high and lower frequency sampling can be conducted simultaneously using a splitter, however this comes at a significant cost of signal-to-noise ratio. This would be particularly harmful during the work presented here which maximizes spin wave propagation distances and thereby minimize power at the detection transducer. 

Fifth, the physical length of the YIG film is much greater than the propagation length $L$ between the transducers. The ends of the film are additionally cut at 45 degree angles. These considerations eliminate end reflections from reaching transducers and thereby impacting ring dynamics.  

Sixth, we use two-element U-shaped antennas as transducers for the excitation and detection of spin waves within the film.  These two-element antennas function as spectral filters by reducing the passband in the AFR. The upper edge of the passband is not effected, however the lower bound now corresponds to the destructive interference condition where the frequency in each of the elements is out of phase by $\pi$.  The use of two-element antennas generally results in a smoother, more linear passband and can limit the spectral distances between ring eigenmodes at higher AFR gains~\cite{Kalinikos2000,Scott2001}. Long lifetime dynamics of SWE solitary waves can more readily be isolated and identified when the range over which the ring gain can be manipulated without introducing spin wave dynamics from vastly differing eigenmode frequencies (and therefore group velocities) into the ring is maximized. This type of filtering is a choice of convenience and experimental expediency, and is not necessary for the generation of dynamics reported in this work. The transducers used in this experiment are 50~$\mathrm{\mu m}$ wide and 2~$\mathrm{mm}$ long with a separation distance between the elements of 60~$\mathrm{\mu m}$. The maximum wavenumber excited by this transducer is given by $k_{\mathrm{max}}=2\pi/\mathrm{d}$ and here $\mathrm{d}$ is the separation between the elements giving $k_{\mathrm{max}}=\mathcal{O}(10^3~\mathrm{rad/cm})$. Note the wavenumber of the lowest loss eigenmode is typically near 100~$\mathrm{rad/cm}$. 

Finally, the variable attenuator and fixed amplifier pair have a linear response in the frequency and power ranges used in the experiment.

\subsection{\label{sec:eandm:SW}Spin Waves in Magnetic Thin Films}

As discussed above, an AFR with these components and design considerations supports the self-generation of backward volume spin wave envelope solitary waves~\cite{Wu2010}.  This is the AFR configuration which has generated a majority of the previously observed dynamics discussed above. BVSWs may be self-generated in an AFR where the propagation medium has been placed in an external magnetic field that is applied parallel to the film, as shown in~\fref{fig:exp_setup}.  BVSWs are so called due to their negative group velocity. They have positive dispersion, $D$, and negative nonlinearity, $N$, coefficients. The opposite signs of dispersion and nonlinearity indicate the effects of one may compensate the effects of the other. This type of nonlinearity is called attractive, or self-focusing, and supports the generation of bright solitary waves, see~\eref{eqn:bright_soliton} below.

The signs and amplitudes of the nonlinearity and dispersion are easily tunable for spin waves propagating in magnetic thin films. This tunability is achieved primarily by rotating the external field relative to the film. For example, forward volume spin waves may be excited by applying the external magnetic field normal to the thin film. This type of excitation also exhibits attractive nonlinearity, but contrary to BVSWs these waves have a positive group velocity, a negative dispersion, and a positive nonlinearity. Surface spin waves may be excited by placing the external field normal to the vector of propagation (or normal and in-plane with the external field shown in~\fref{fig:exp_setup}). Surface spin waves have repulsive nonlinearity, with both the dispersion and nonlinearity coefficients being negative. This type of nonlinearity supports the propagation of dark solitary waves, see~\eref{eqn:dark_soliton} below. The sign and magnitude of dispersion may also be adjusted by choosing operating points on either side of dispersion gaps for pinned surface waves~\cite{Wu2004}. Again, only the BVSW regime is considered in this work, but interested readers are directed to these references for further information~\cite{Wu2010,Stancil2009,Chen1994,Boardman1995,Wigen1994}. We note there is also ongoing research into exciting multiple spin wave regimes at once~\cite{Bang2018}.

Nonlinearity and dispersion coefficients for spin waves in nonlinear thin films are defined in terms of the dispersion relation, $\omega(k)$ as
\begin{equation*}
D = \frac{\partial^2\omega(k)}{\partial k^2},
\end{equation*}
\begin{equation*}
N = \frac{\partial\omega(k)}{\partial|u|^2},
\end{equation*}
where $u$ is a dimensionless spin wave amplitude. These expressions for nonlinearity and dispersion are results of deriving the governing equation of spin waves in thin films via a slowly varying envelope approximation on the nonlinear dispersion relation, $f( \omega , k , |u|)$~\cite{Stancil2009}. The resulting equation is,
\begin{equation}\label{eqn:model:GLNLS}
i\frac{\partial u}{\partial t}=\left[-\frac{D}{2}\frac{\partial^2}{\partial x^2} + N\vert u\vert^2+ V(x,t,u)\right]u
\end{equation}
and when the potential $V$ is set to zero we have a dimensionless form of the nonlinear Schr\"odinger equation. 

Higher order nonlinearity may be naturally included by keeping more terms of the expansion of the dispersion relation.  Losses are introduced phenomenologically. If cubic and quintic terms are kept then one arrives at the CQCGL equation where the potential is of the form
\begin{equation}\label{eqn:model:GLNLS2}
V(x,t,u)= iL +iC\vert u\vert^2 +(S+iQ)\vert u\vert^4,
\end{equation}
here $L$, $C$ and $Q$ are linear, cubic and quintic loss coefficients and S is a quintic nonlinearity coefficient.  All parameters are taken to be real with i's explicitly stated, except the spin wave amplitude $u$ which is a complex scalar field.  The inclusion of at least quintic terms is necessary to model saturable losses that are experimentally observed in self-generated spin waves which evolve under either three or four-wave mixing~\cite{Scott2004,Hagerstrom2009}. Saturable losses are also studied in dissipative optical solitons~\cite{Ablowitz2008,Akhmediev2008}. Nonlinear Schr\"odinger governing equations for spin waves in nonlinear thin films may also be rigorously derived through a Hamiltonian formalism~\cite{Krivosik2012}.

When four-wave mixing, discussed below, is the dominant source of nonlinearity and losses are negligible (or $V$=0) the ground state solitary solution to the nonlinear Schr\"odinger equation may be reached via an inverse scattering transform~\cite{Ablowitz2003}. For the case of attractive nonlinearity ($ND<0$) the ground state solution is the standard bright soliton given by
\begin{equation}\label{eqn:bright_soliton}
u(z,t) = \mu_0\, \mathrm{sech}\left[\mu_0\sqrt{\frac{-N}{D}}\left(z-v_\mathrm{s} t\right)\right],
\end{equation}
where the phase is given by
\begin{eqnarray*}
\phi(z,t)&=&\exp{i[k_\mathrm{s} z - \omega_\mathrm{s} t]} \\
v_s&=&v_g+k_sD \\
\omega_s &=&  v_\mathrm{g} k_\mathrm{s} + \frac{1}{2}D k_\mathrm{s}^2 + \frac{1}{2}N \mu_\mathrm{0}^2
\end{eqnarray*}
with $u(z,t)=|u(z,t)|\exp(i\phi(z,t))$.  We highlight that for the bright soliton the phase is constant over the soliton peak and there is no phase difference at $(z-v_\mathrm{s})\rightarrow\pm\infty$. The hyperbolic secant shape and the phase characteristics across the peak are common identifiers of bright solitary waves. 

Dark solitary waves are the ground state solution for repulsive nonlinearity ($ND>0$) and are given by
\begin{equation}\label{eqn:dark_soliton}
|u(z,t)| =\mu_\mathrm{0} \left[1-\mathrm{sech}^2\left(\mu_\mathrm{0}\sqrt{\frac{N}{D}}\left[z-v_\mathrm{s} t\right]\right)\right],
\end{equation}
and the phase across the soliton peak is
\begin{equation}\label{eqn:dark_soliton_phase}
\phi(z,t) = \mathrm{arctan}\left[\mathrm{tanh}\left(\mu_0 \sqrt{\frac{N}{D}}\left[z-v_g t\right]\right)\right]
\end{equation}
Dark solitons are notches in the amplitude of a continuous wave background and possess a $\pi$ shift across their dips. If the final depth of the notch does not reach zero then the soliton is called grey and the phase shift is less than $\pi$, somewhat modifying~\cite{kivshar1998dark} the form given in equations~\ref{eqn:dark_soliton} and \ref{eqn:dark_soliton_phase}. As with bright solitary waves the hyperbolic secant dip and the phase characteristics across the dip are common identifying features of dark and grey solitary waves. Note that there is a phase difference at $(z-v_\mathrm{s})\rightarrow\pm\infty$ for dark and grey solitary waves. This means that if these waves are traveling within a ring there must always be a minimum of two to satisfy phase continuity conditions unless there is a background flow of $-\pi$ to cancel the phase jump~\cite{kanamoto2008topological}. Typically bright (dark) solitary waves do not exist outside of the attractive (repulsive) nonlinearity which supports them as a solution to the nonlinear Schr\"odinger equation. Any exceptions to this suggest the influence of higher order or other external effects~\cite{scott2005}.As mentioned earlier, the sign of dispersion can also be control by exciting spin waves on either side of a a dipole gap in a film with pinned surface waves, thus modifying of the effective nonlinearity, $ND$, and enabling the excitation of dark solitary waves~\cite{Wu2004}. Similarly an external periodic potential can also be applied to generate negative mass and thus flip the sign of dispersion. It has also been demonstrated that dark solitary waves may be observed in spin wave systems with attractive nonlinearity if one injects carefully tuned carrier wave signals~\cite{scott2005}. However, implementing a periodic potential is impractical in many contexts including fiber optics, and the long lifetime dynamics presented here are not observable when carrier wave signals are fed into a short film with no active feedback.

There are two types of nonlinearity which affect spin waves in YIG waveguides: the nonconservative three-wave splitting or confluence processes, and the conservative four-wave mixing process~\cite{Stancil2009,Wu2010,Wigen1994} (known as three and four-magnon scattering in quantum mechanics). The bright and dark solitary wave solutions above and the solitary wave dynamical behaviors presented below are principally products of four-wave mixing.  During the self-generation of spin waves in an AFR if two or more eigenmodes are circulating the ring with sufficient amplitude they interact via four-wave mixing to generate a third mode ($\omega_3$) with the following frequency and wavenumber
\begin{eqnarray*}
2\omega_1 &=& \omega_2+\omega_3\,,\\
2k_1 &=& k_2+k_3.
\end{eqnarray*}

\begin{figure}
\includegraphics[width=.5\textwidth]{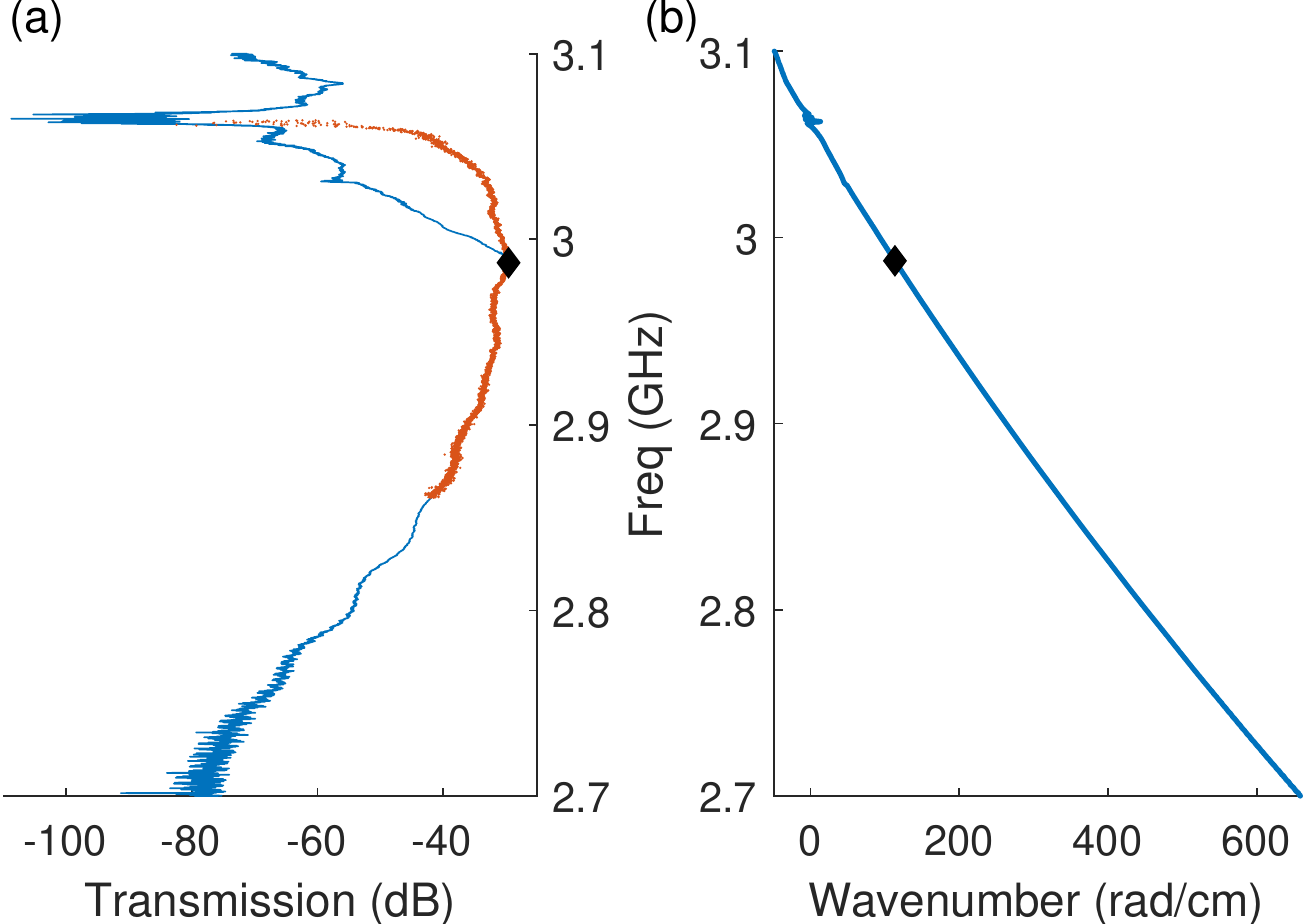}
\caption{\label{fig:transmission}\textit{Yttrium iron garnet transmission profiles} (a) Transmission loss and (b) wavenumber vs frequency profiles for a transducer-YIG thin film-transducer structure with parallel external magnetization. The red and blue curves in (a) were measured at input powers $P_\mathrm{in}$=-24~dBm and 4~dBm, respectively. Additional transmission losses at high power indicate three-wave mixing. The lowest loss ring eigenmode is shown as a black diamond at 2.987~GHz and corresponds to a wavenumber of $k\approx114$~rad/cm.}
\end{figure}

As power is further increased additional modes will be generated through this four-wave process, ultimately resulting in a uniform frequency comb with spacing $f_\mathrm{s}=|\omega_1-\omega_2|$. These equispaced frequency combs are a spectral signature of bright and dark solitary waves and importantly cannot be generated simply through exciting additional ring modes owing to the spin wave nonlinear dispersion relation. An experimentally measured dispersion curve can be seen in~\fref{fig:transmission}(b), where the $k=0$ point corresponds to the uniform mode or the ferromagnetic resonance in the YIG thin film. The lowest loss eigenmode at $2.987~\mathrm{GHz}$ is shown as a black diamond and corresponds to a wavenumber of $k\approx114$~cm/ns. This data was recorded at an external field strength of $496~\mathrm{Oe}$.

The three-wave splitting and confluence processes involve the splitting of one mode ($\omega_0$) into two new half-frequency modes, or the combination of two half-frequency modes into a new mode ($\omega_3$) and is inherently nonconservative. The splitting and confluence processes are given by
\begin{eqnarray*}
\omega_0 &=& \omega_1+\omega_2,\\
k_0 &=& k_1+k_2,
\end{eqnarray*}
and
\begin{eqnarray*}
\omega_1+\omega_2 &=& \omega_3,\\
k_1+k_2 &=& k_3,
\end{eqnarray*}
respectfully. For BVSWs the three-wave process can be viewed as a source of nonlinear loss where energy leaves the ring system by the splitting of self-generated ring eigenmodes, with wavenumbers of $\mathcal{O}(10~\mathrm{rad/cm})$, into two half frequency modes with wavenumbers of $\mathcal{O}(10^5~\mathrm{rad/cm})$~\cite{Hagerstrom2009,Wu2010}.  These half frequency modes posses wavenumbers well beyond the maximum wavenumber capable of being excited or detected by the transducers used in this experiment, as explained above. Therefore the half modes do not circulate within the ring. Without the active feedback these modes decay into heat and there is no associated confluence process to introduce power back into the higher frequency modes. Note, the major linear losses in YIG thin films result from magnon-phonon scattering, an intrinsic relaxation process, and from two-magnon scattering, an extrinsic relaxation process associated with film defects~\cite{Stancil2009}.

An experimental example of transmission loss through the transducer-YIG thin film-transducer structure (not including the feedback ring) is shown in~\fref{fig:transmission}(a) with a high power (4~dBm) transmission loss curve shown in blue and a low power (-24~dBm) transmission loss shown in red.  The loss caused by three-wave mixing is evident in the high power data above 2.99$~\mathrm{GHz}$. This closely matches with theory that predicts the lower edge of the passband for BVSWs at 1.50$~\mathrm{Ghz}$. Below 3$~\mathrm{Ghz}$ the three-wave splitting process is forbidden by conservation law. Note that at sufficiently high external field strengths, no half modes remain within the BVSW passband and the three-wave processes are entirely forbidden. For BVSWs this occurs when $\omega(0)=3.27~\mathrm{GHz}$ or at an external field strength of about $600~\mathrm{Oe}$~\cite{Chen1994}. The general shape of the low power transmission loss plot in~\fref{fig:transmission}(a) is typical.  The transducers are generally less efficient for high-k modes, but this is directly counteracted by the increase in group velocity of BVSWs at higher frequencies. An increase in the group velocity decreases the round trip time, and thereby the overall losses. The result is a fairly flat transmission loss curve spanning approximately 400 wavenumbers with a minimum in the middle.

At lower external fields the loss from three-wave splitting will span more of the experimental passband. By $350~\mathrm{Oe}$ the full passband will experience additional losses from three-wave splitting. In practice this results in a saturation in the power-out vs power-in response, or the addition of higher order loss. Note that it is well documented that BVSWs in AFRs experience a saturation in power-out vs power-in even when three-wave processes are forbidden, so three-wave mixing can be viewed as an additional, and stronger, source of nonlinear losses~\cite{Hagerstrom2009,Scott2004}.

The data presented here were gathered at four distinct operating frequencies across different choices of transducer separation. Spin waves were self-generated at 2.5~$\mathrm{GHz}$, where the entire passband is subjected to additional nonlinear losses, at 3.0~$\mathrm{GHz}$, where part of the passband experiences three-wave splitting such as in~\fref{fig:transmission}(a), and at 3.5~$\mathrm{GHz}$ and 4.5~$\mathrm{GHz}$ where three-wave splitting is explicitly forbidden. In practice, actual operating points vary from these targets as external field strength and the location of the film on the transducers is varied to isolate clean dynamics. Higher frequency excitations were not explored in order to maximize the impact of ring losses on ring dynamics. We highlight that this work presents the first observations of self-generationed dark solitary waves under attractive nonlinearity, and the first example of long life time complex dynamical behaviors in dark solitary waves.

\section{\label{sec:bre}Periodic Breathers}
Periodic breathing is a stable dynamical pattern characterized by the smooth, periodic amplitude modulation of SWE solitary waves matched with a simultaneous relocation of the wave positions within the feedback ring. Three features define this behavior. (1) The amplitude and direction of the location shift are constant. This amount is always proportional the number of solitary waves in the ring. For $N$ solitary waves a full period is defined by $(N+1)$ breaths with positional shifts of $2\pi/(N+1)$~rad. (2) A minimum of one solitary wave must have its amplitude drop to the background during a full breathing period. This is necessary to draw a qualitative distinction between breathers and modulating solitary wave trains where higher order solitons exhibit periodic or chaotic amplitude modulation but without relocation~\cite{Wu2010,Wang2011}. (3) The shift is not spontaneous but rather a smooth function of round trips which remains consistent across all observed breathing periods. To summarize, in scaled units, a periodic breathing SWE solitary wave oscillates between a magnitude of zero and one while relocating predictably within the ring.

Periodic breathing for both bright and dark solitary waves are presented for two typical examples of the behavior. (1) In periodic breathing solitary waves relocate at a single fixed period. (2) In multi-periodic breathing the solitary waves breathe at two or more frequencies. We highlight, as mentioned in \sref{sec:eandm}, that all data presented here were recorded on an AFR with a parallel external magnetic field. This film geometry enables the excitation of BVSWs with an attractive nonlinearity. The dark solitary wave dynamics reported here occur in a geometry which traditionally supports the generation of bright solitary waves via the compensation of dispersion with attractive nonlinearity and loss. The only experimental parameters changed between observations are external field strength, AFR gain and transducer separation. None of these parameters explicitly alter the sign of cubic nonlinearity operating within the film.

\begin{figure}
\subfloat[][\label{fig:breath_1a}]{\includegraphics[width=1\columnwidth]{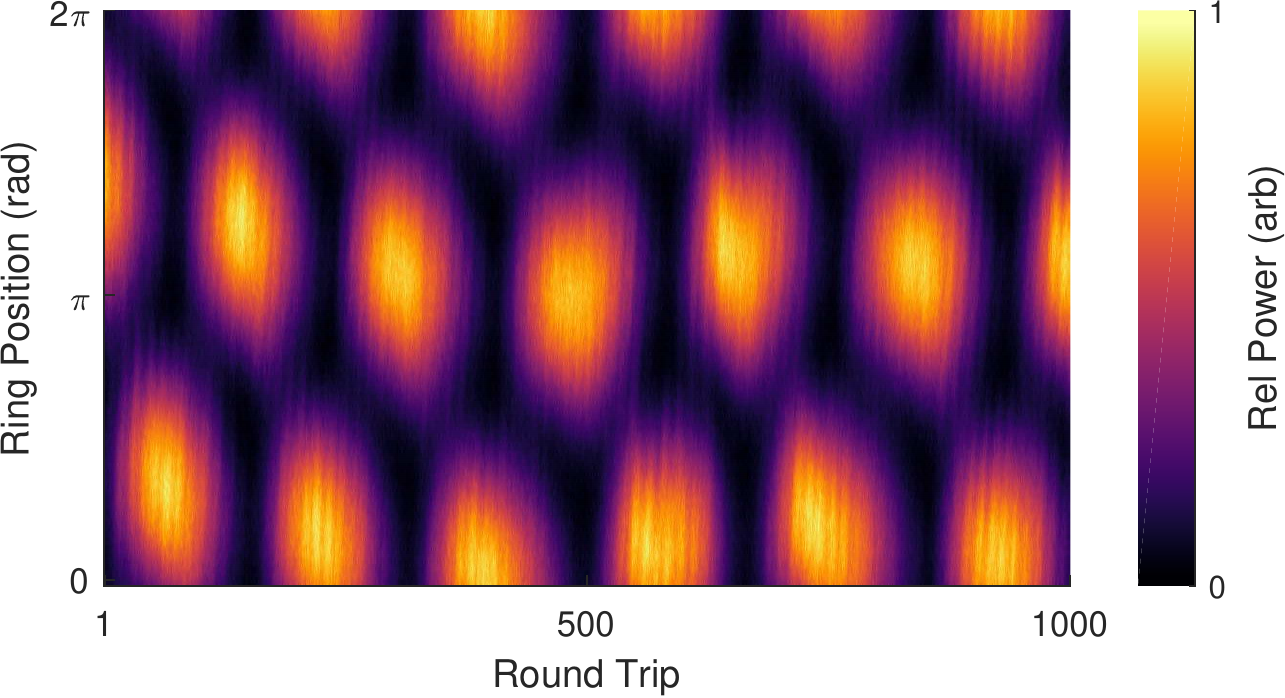}}\\
\subfloat[][\label{fig:breath_1b}]{\includegraphics[width=.49\columnwidth]{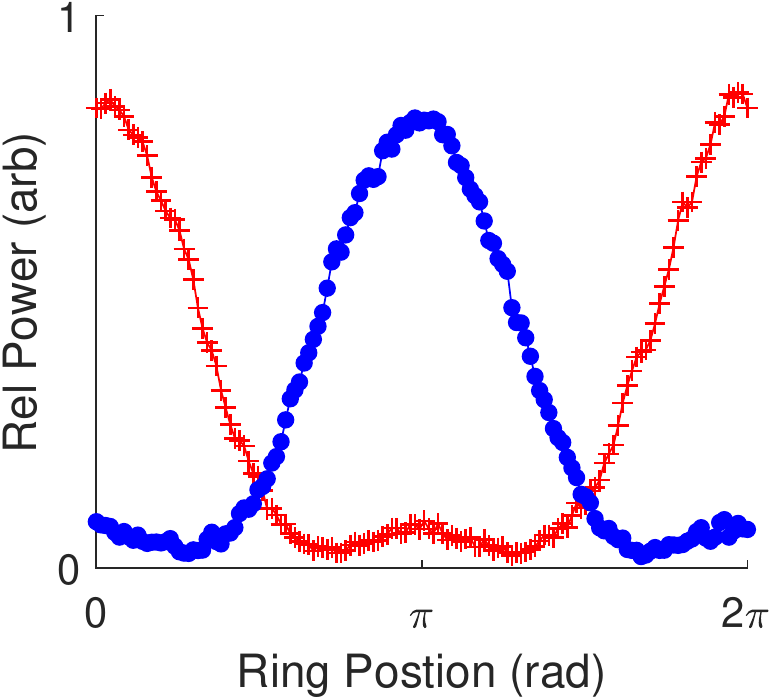}}~
\subfloat[][\label{fig:breath_1c}]{\includegraphics[width=.49\columnwidth]{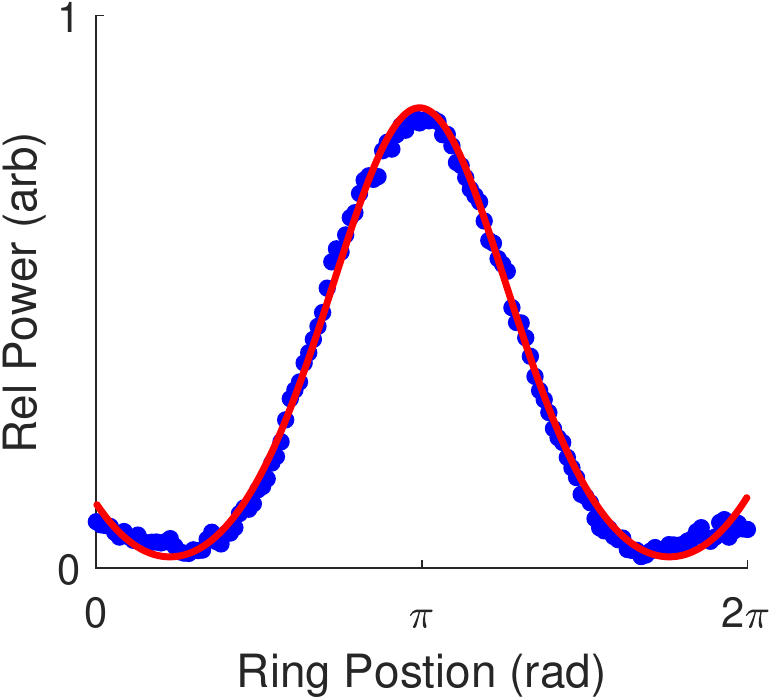}}
\caption{\label{fig:breath_1}\textit{Bright solitary wave periodic breathing.} The stable periodic breathing of one bright solitary wave from the center to the edge of the ring ($\pi$ shift) is shown in the spatiotemporal plot (a) The bright solitary wave nature of the dynamic is shown in the scaled power vs. ring position plots (b) and (c). Plot (b) shows round trips with the peak in the center and edge of the ring as blue dots and red crosses, respectively.  A fit to a Jacobi elliptic CN, a periodic generalization of the sech function~\cite{abramowitz1964}, is shown in plot (c) with the peak centered in the ring as blue dots and the fit as a red line. Actual data is binned for clarity, and lines in (b) are a guide for the eye.}
\end{figure}

\subsection{\label{sec:bre:clean:bright}Bright Solitary Wave Periodic Breathing}

Periodic breathing is epitomized by a single, bright solitary wave modulating at one, fixed frequency. In this representative case the behavior is easily observable and the consistency of the wave's predictable relocation within the ring is apparent. A typical example, recorded through a diode at an external field strength of $311~\mathrm{Oe}$ with a $2.39~\mathrm{GHz}$ carrier wave, is shown in~\fref{fig:breath_1}. The bright solitary wave breather is readily identifiable through its reconstructed spatiotemporal amplitude plot, given in~\fref{fig:breath_1a}, where each vertical slice is a single round trip and we move through time (in the group velocity frame) from left to right. Scaled spin wave power is shown as shading. Here we see a bright solitary wave (with a scaled amplitude of one) centered in the ring which smoothly decreases in amplitude to zero prior to undergoing a $\pi$ shift to the edge of the ring. The wave then breathes back to the center, completing a single period. The breathing process then repeats. The bright solitary wave profiles for a single round trip at the maximum peak amplitude of a center (dotted-blue) and edge (crossed-red) breath are plotted in~\fref{fig:breath_1b}. Actual data is binned for clarity and lines are a guide for the eye. We highlight that the plots in~\fref{fig:breath_1} correspond to solitary waves circulating within the AFR but have been shifted into a frame co-moving with the waves at their group velocity. This type of reconstructed spatiotemporal plot will be used extensively throughout this work to highlight long lifetime solitary wave dynamics. Similarly through this work, unless stated otherwise, actual data is plotted and lines are a guide for the eye. 

\begin{figure}
\includegraphics[width=1\columnwidth]{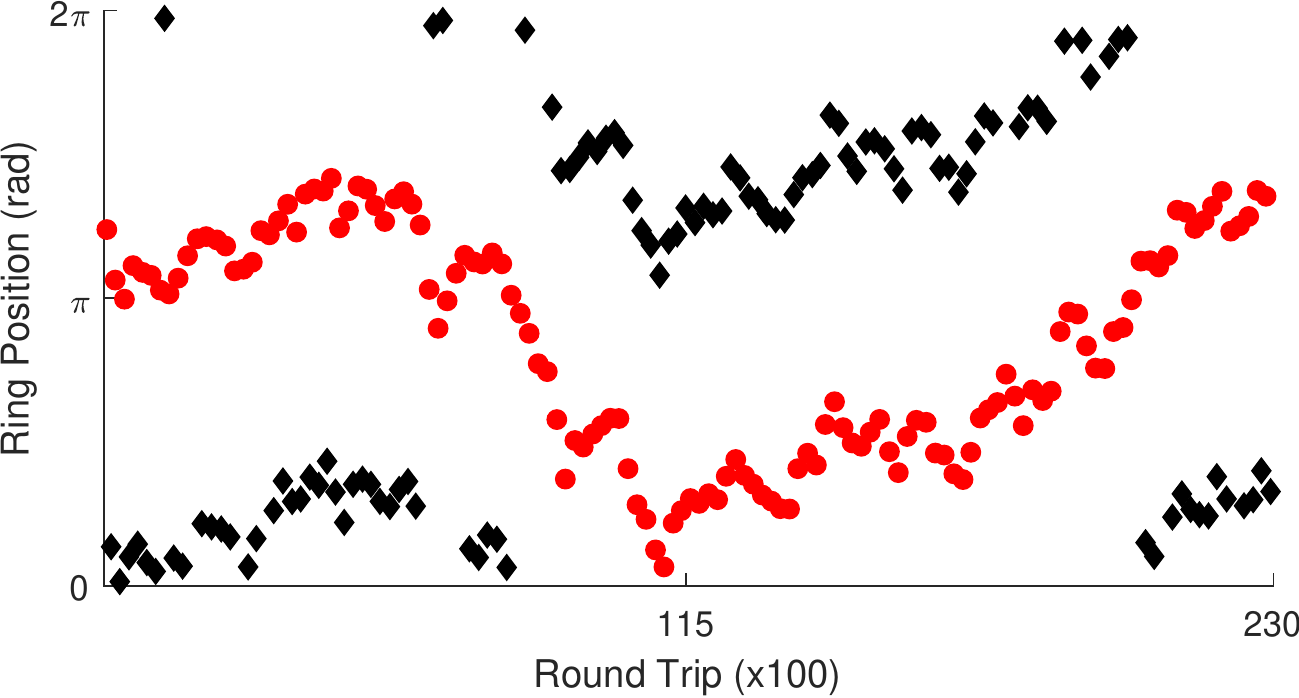}
\caption{\label{fig:breath_3}\textit{Bright solitary wave breathing.} Peak locations for central (red dots) and edge (black diamonds) breaths for a bright solitary wave undergoing stable periodic breathing.}
\end{figure}

A round trip time of $353.625~\mathrm{ns}$ was identified visually by minimizing the change in the central breathing location over all observed breathing periods. A total of $8.2~\mathrm{ms}$ of data was collected at $4~\mathrm{Gsamples/s}$ yielding over 23,000 round trips of data. 1000 round trips are shown in~\fref{fig:breath_1a}. This data was collected through a diode so, as discussed in~\sref{sec:eandm}, only the spin wave envelope was accurately captured. A peak finding algorithm was used to isolate 261 individual peak locations, or 130 full breathing periods. These peak locations are visualized in~\fref{fig:breath_3} with the central peaks as red dots and the edge peaks as black diamonds. These locations were then used to generate profiles across round trips (temporal) and across ring position (spatial) for each of the peaks. A breathing period of 175.98 $\pm$ 17.68 round trips was identified from these statistics, suggesting 132 breathing periods were recorded. This is consistent with our peak finding algorithm where we ignored the first and last two breathing periods to ensure full temporal and spatial profiles could be isolated.

Some jitter in the group velocity is observed across the 23,000 round trips. The peaks can be seen drifting around the ring relative to their original positions. Using all 261 isolated peaks the change in group velocity is found to be -7.95x$10^{-7} \pm 3.75x\times 10^{-4}$ cm/ns per breathing period, or -1.28x$10^{-13} \pm 6.03 \times 10^{-9}$ cm/$\mathrm{ns}^2$. The group velocity jitter is determined to be normally distributed about this near-zero mean with an Anderson-Darling statistic of 0.270 (normality would be rejected at 0.750). Given a measured transducer separation of $1.233 \pm 0.040~\mathrm{cm}$ and a round trip time of $353.625~\mathrm{ns}$ we can estimate the the group velocity as $-3.487 \times 10^{-3} \pm 1.1 \times 10^{-4}$ cm/ns. The jitter, then, amounts to at most an order 10\% change in group velocity about the mean, or a total shift in ring location of less than 10\%, over a single breathing period.

\begin{figure}
\subfloat[][\label{fig:breath_2a}]{\includegraphics[width=.49\columnwidth]{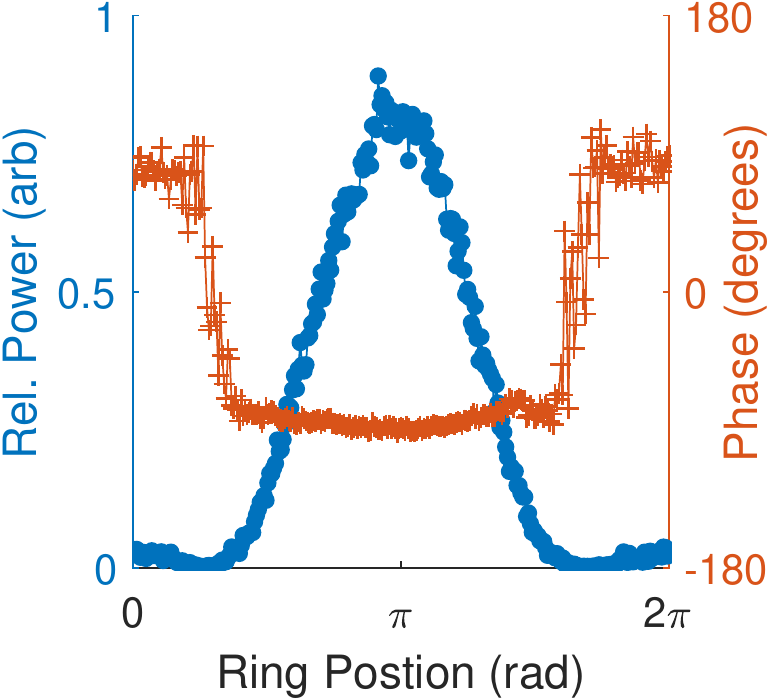}}~
\subfloat[][\label{fig:breath_2b}]{\includegraphics[width=.49\columnwidth]{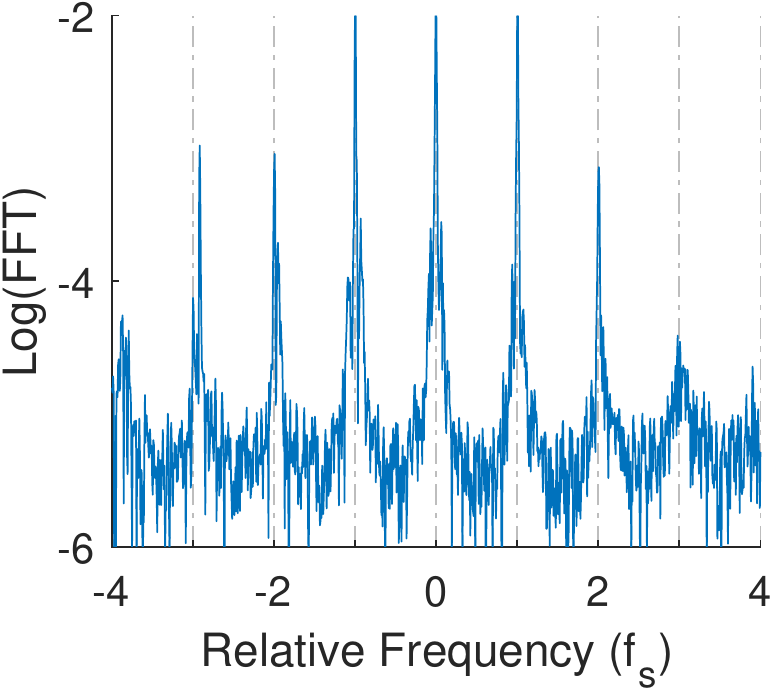}}
\caption{\label{fig:breath_2}\textit{Bright solitary wave breathing.} Reconstructed phase data which demonstrates the solitonic nature of the stable periodic breathing wave. (a) Ring power and reconstructed phase data in blue dots and red crosses, respectively. A flat phase across the peak profile is a defining characteristic of bright solitary waves. (b) Power spectrum around the principal carrier frequency and scaled by the comb spacing, $f_{\mathrm{s}}$. An equi-spaced frequency comb generated via four-wave mixing is another defining feature of solitary waves.}
\end{figure}

The solitonic nature of the underlying wave is readily verified by the fit shown in~\fref{fig:breath_1c} for a single of the 262 identified peaks (chosen at random). Here actual data is plotted and the red line is a fit to the generalized bright soliton solution to the NLS given in~\fref{eqn:bright_soliton}. Each of the 261 identified peaks was fit using nonlinear least squares in order to generate goodness-of-fit statistics. The fits yielded a mean reduced $\chi^2$ of 2.65$\pm$2.22, a mean adjusted $\mathrm{r}^2$ of 0.994$\pm$0.008 and with a Jacobi-elliptic parameter~\cite{abramowitz1964}, $m$, of 0.6862$\pm$0.18. A value of $m$ this close to 1 indicates the solitary wave is more hyperbolic than sinusoidal and may be qualitatively described as a hyperbolic peak profile with tails which do not quite go all the way to zero. Fits were made to purely solitonic and hyperbolic profiles with reduced $\chi^2$ 4 and 7 times larger than the Jacobi-elliptic fit, respectively.

Phase data, reconstructed from data collected at $40~\mathrm{Gsamples/s}$ without a diode, also confirms the solitonic features of the data.  A single round trip is shown in~\fref{fig:breath_2a} where the spin wave envelope is plotted as blue dots and a reconstructed phase is plotted as red crosses. The flat phase across the peak profile is a defining characteristic of bright soliton solutions to the NLS, including the general solution given previously in~\eref{eqn:bright_soliton}. The dynamics are additionally verified as being dominated by four-wave mixing by the power spectrum given in~\fref{fig:breath_2b}. A comb-like, evenly spaced spectrum with 3 identifiable side peaks around the central carrier frequency is observed with a fixed separation of $f_\mathrm{s}=0.28292~\mathrm{MHz}$. This equispaced frequency comb is the defining characteristic of four-wave mixing, generated by the conservative equations given in~\sref{sec:eandm}. This frequency difference also corresponds to a round trip time of $353.456~\mathrm{ns}$, which agrees with the round trip time identified visually from~\fref{fig:breath_3}. The dispersion and nonlinearity coefficients were estimated from fitting recorded dispersion curves and found to be 1.49x$10^{-6}~\mathrm{cm^2rad/ns}$ and $-6.27~\mathrm{rad/ns}$, respectively.

\begin{figure}
\subfloat[][$P_\mathrm{RMS}=1.0$ (arb)\label{fig:breath_4a}]{\includegraphics[width=0.49\columnwidth]{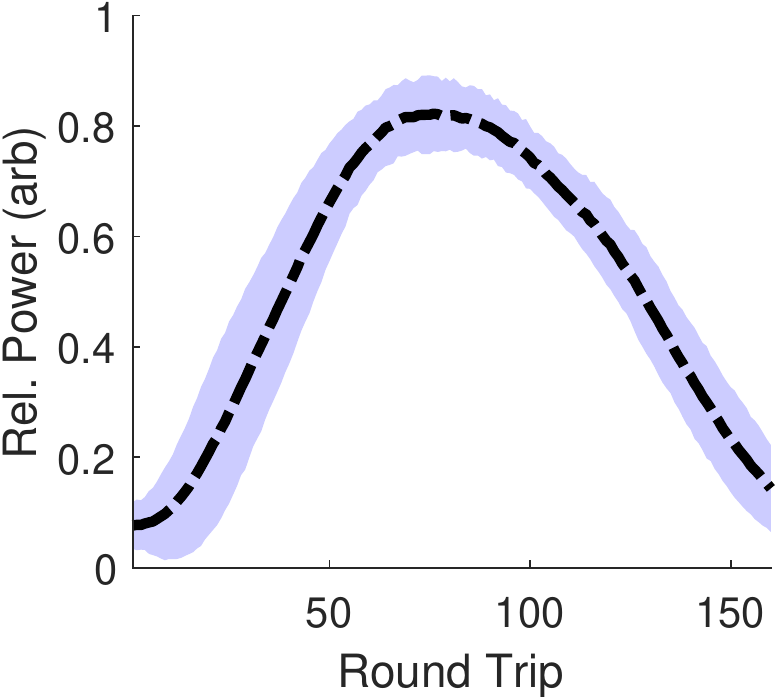}}~
\subfloat[][$P_\mathrm{RMS}=1.02$ (arb)\label{fig:breath_4b}]{\includegraphics[width=0.49\columnwidth]{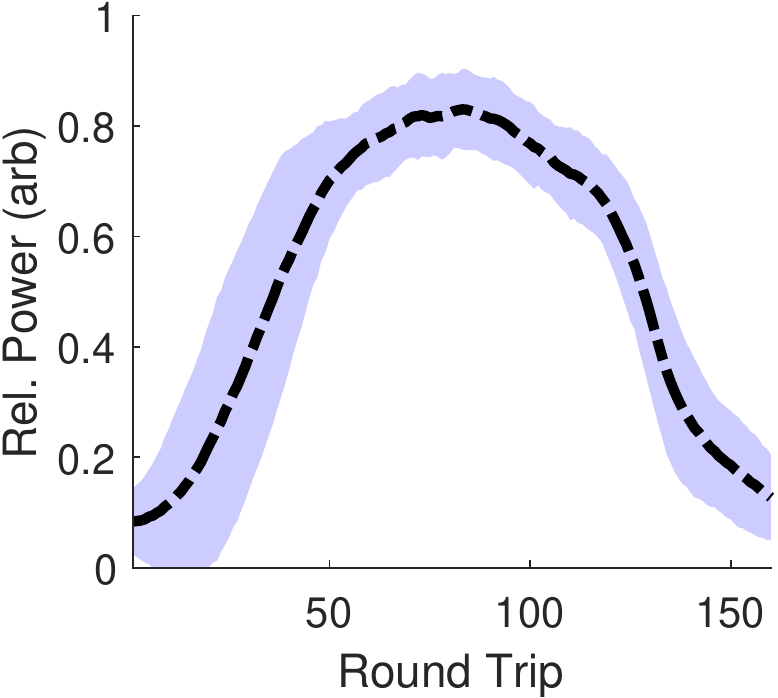}}\\
\subfloat[][$P_\mathrm{RMS}=1.12$ (arb)\label{fig:breath_4c}]{\includegraphics[width=0.49\columnwidth]{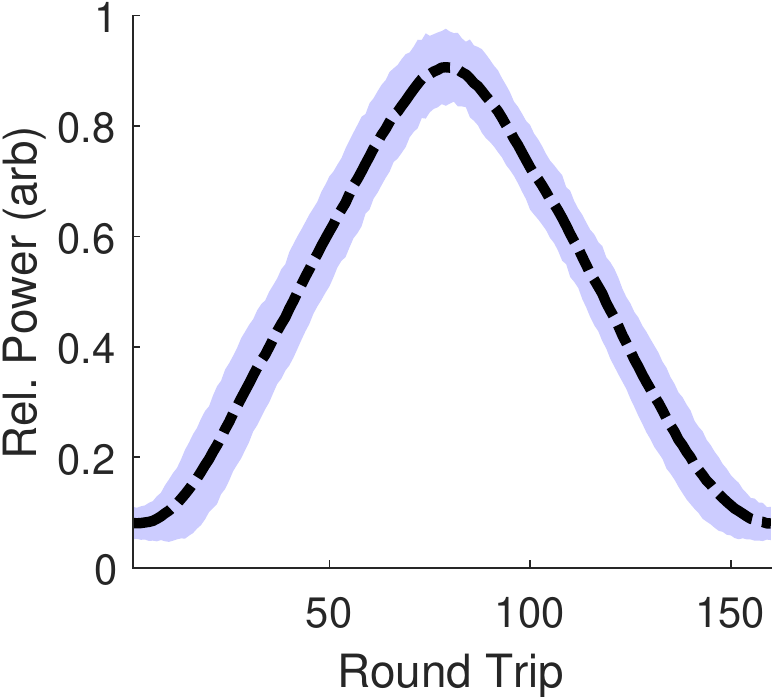}}
\caption{\label{fig:breath_4}\textit{Bright solitary wave periodic breathing.} The breathing profile of a single peak across round trips. Mean of all 262 observed breaths shown as a dotted black line, with a confidence interval constructed from 2 times the standard deviation shown as light blue. Active feedback ring gain is increased from (a) to (c) with the mean RMS power of the observed peaks normalized to low power, (a), given as individual labels. As ring gain is increased the breathing transition becomes sharper, with less time spent at high amplitude.}
\end{figure}

It is important to note that an external field of $311~\mathrm{Oe}$ is not large enough to drive the AFR to a power which forbids the nonconservative three-wave splitting process from occurring within the YIG film. As discussed in~\sref{sec:eandm} for our AFR geometry, which supports the self-generation of BVSWs, the three-wave mixing process generates half modes which are not observable with the transducers used in this experiment. However, while a stable solitary wave is circulating within the ring, four-wave mixing must remain the dominant dynamical process. Three-wave mixing is therefore considered an additional nonlinear loss. A fully defined bright soliton typically propagates in rings with powers high enough to generate 6 or more four-wave mixing side peaks. The absence of these additional frequency modes here, as well as the lower than unity Jacobi-elliptic parameter, indicate that this solitary wave breathing dynamic occurs at lower ring gains than those which support bright soliton wave trains and where higher order losses impact dynamics. This matches the numerical predictions for regimes which support solitary wave periodic breathing where cubic losses were the highest relative loss~\cite{anderson2014complex}.

A total of three distinct bright single periodic breathers were observed experimentally at increasing ring gain. All were observed at an external saturation field magnitude of $311~\mathrm{Oe}$.  Each breather shared their core characteristics including the following. (1) Their spatial peak profiles fit to a Jacobi-elliptic with a parameter, $m$, within two standard deviations of 0.69. (2) Half-way through relocation, when both peaks are of equal amplitude, their spatial profile is purely sinusoidal. (3) Parameters such as round trip time, breathing period and group velocity jitter are all within 1\% of one another.

As ring gain is increased the temporal profiles of the breathers evolve from purely sinusoidal to being best described by a sharper peak, given by the generic NLS Jacobi-elliptic solution with a parameter of $m=0.22\pm0.11$. The temporal profiles are shown in~\fref{fig:breath_4} with scaled ring RMS round trip power at maximum peak height increasing from $1.00\pm0.04$ in~\fref{fig:breath_4b} to $1.12\pm0.05$ in~\fref{fig:breath_4c}. Here the mean peak is given by a dotted line and a confidence interval is shown as light blue shading. The interval is constructed by $\pm2\sigma$ where $\sigma$ is the standard deviation from all isolated peaks. This sharpening of the temporal profile at higher ring gains is consistent with our hypothesis that three-wave mixing acts as an additional nonlinear loss, impacting the evolution of peak amplitude over long times.

\subsection{\label{sec:bre:clean:dark}Dark Solitary Wave Periodic Breathing}

A periodically breathing dark solitary wave was observed at an external field strength of $446~\mathrm{Oe}$ with a carrier frequency of $2.94~\mathrm{GHz}$. We again highlight that this dynamic was observed on the same film, on the same day, using the same procedure and experimental setup as was used to generate the bright solitary wave breathers presented in the previous section,~\sref{sec:bre:clean:bright}. The sole adjustment made between the observation of bright solitary waves and the observation of dark solitary waves was external field strength.

We reiterate that these waves are were not created by injecting carrier waves or from exploiting surface pinning and are the first reported self-generation of dark solitary waves in attractive nonlineary~\cite{scott2005,Wu2004}.

\begin{figure}
\subfloat[][\label{fig:breath_5a}]{\includegraphics[width=1\columnwidth]{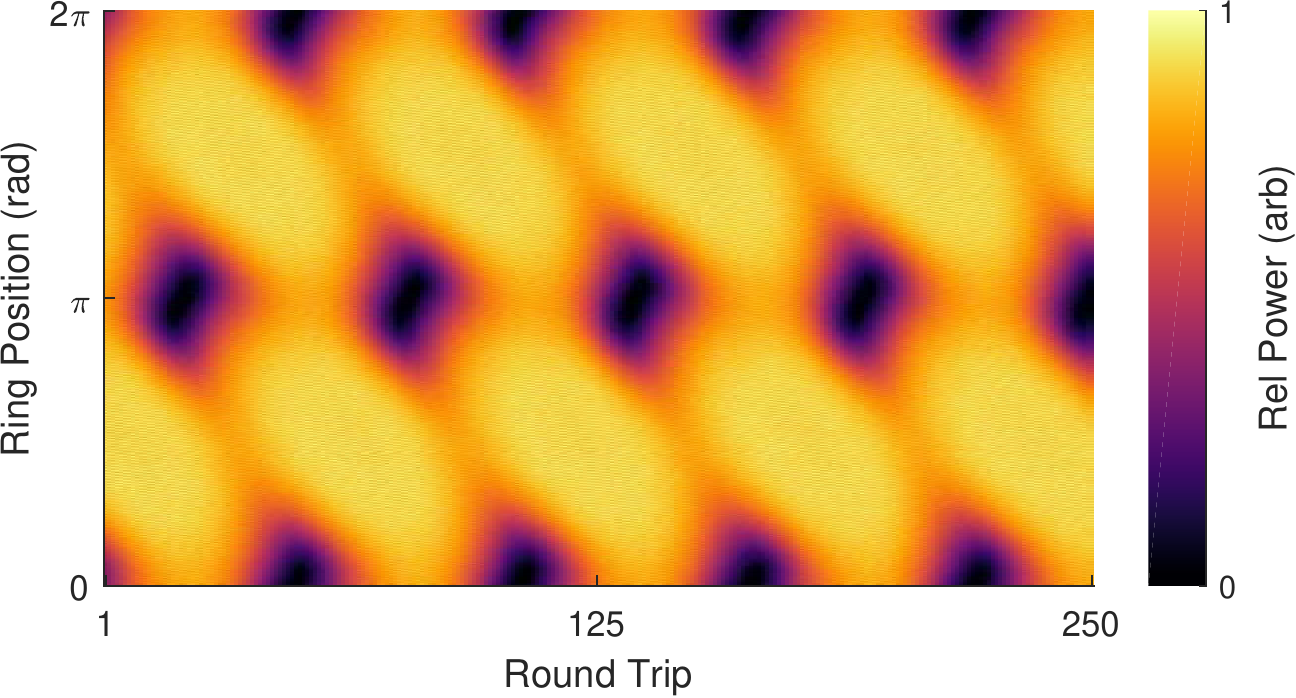}}\\
\subfloat[][\label{fig:breath_5b}]{\includegraphics[width=.49\columnwidth]{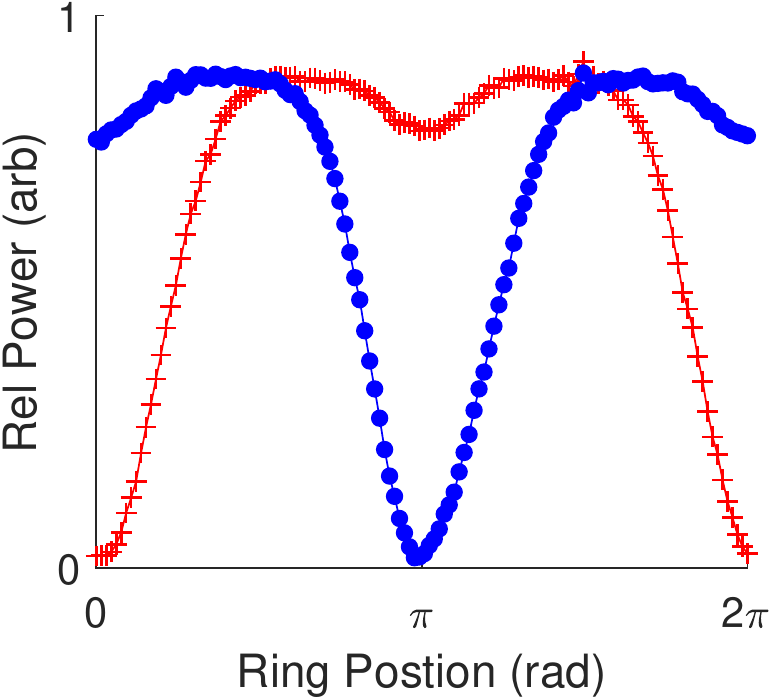}}~
\subfloat[][\label{fig:breath_5c}]{\includegraphics[width=.49\columnwidth]{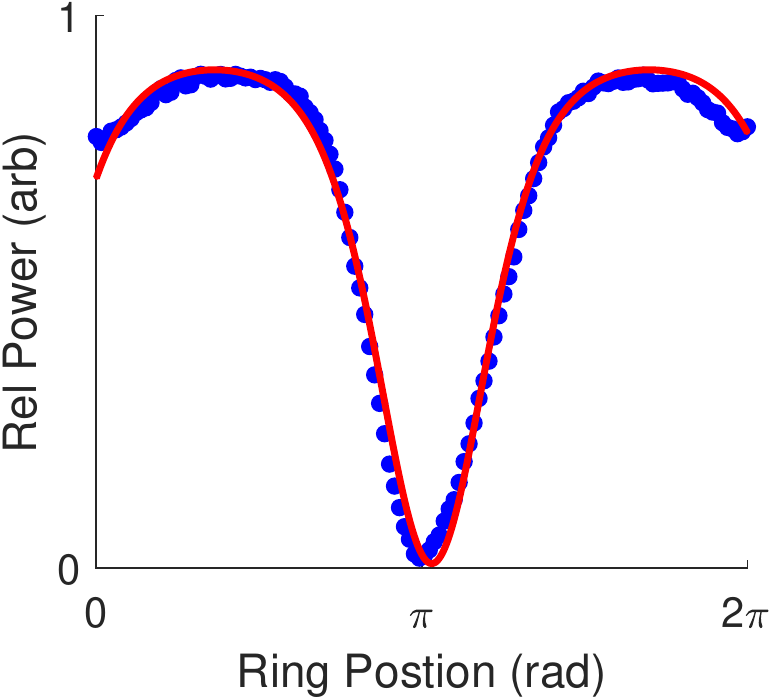}}
\caption{\label{fig:breath_5}\textit{Dark solitary wave periodic breathing.} A single dark solitary wave breathing at a fixed frequency from the center of the feedback ring to the edge. (a) A spatiotemporal plot of the spin wave power. (b) A single round trip of the dark solitary wave in the center of the ring (dotted blue) and at the edge of the ring a half breath later (crossed red). (c) A fit of a single central round trip to the generalized dark solitary Jacobi-elliptic solution to the NLS.}
\end{figure}

A dark solitary wave periodic breather is shown in~\fref{fig:breath_5} where the breathing notches are evident in the reconstructed spatiotemporal~\fref{fig:breath_5a} (low amplitude in black). The dark solitary wave smoothly modulates between a maximum (relative to the background) amplitude of one to zero while undergoing a $\pi$ positional shift within the ring. This meets the qualitative requirements of a periodic breather discussed early in section~\ref{sec:bre}.

A round trip time of 329.34~ns is identified by minimizing the variation in center peak position across all observed 24,000 round trips, or 8.2~ms of data. A peak-fitting algorithm was used to isolate 870 individual peaks and build statistics.  A breathing period of 57.18 $\pm$5.01 round trips was determined from these data.  A group velocity of $-3.74 \times 10^{-3}\pm 3.07 \times 10^{-5}$~cm/ns was calculated using this round trip and the measured transducer separation of $1.233 \pm 0.040~\mathrm{cm}$. No notable jitter of the group velocity was observed. A dispersion coefficient of $1.73 \times 10^{-6}$~cm$^2$rad/ns and a nonlinearity coefficient of -7.23~rad/ns were measured by fitting to observed dispersion curves.

\begin{figure}
\subfloat[][\label{fig:breath_6a}]{\includegraphics[width=.49\columnwidth]{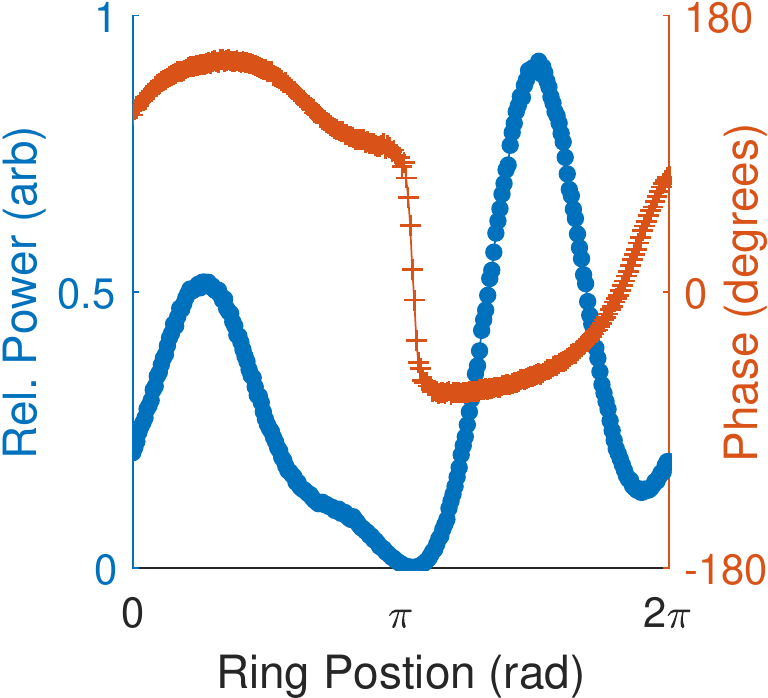}}~
\subfloat[][\label{fig:breath_6b}]{\includegraphics[width=.49\columnwidth]{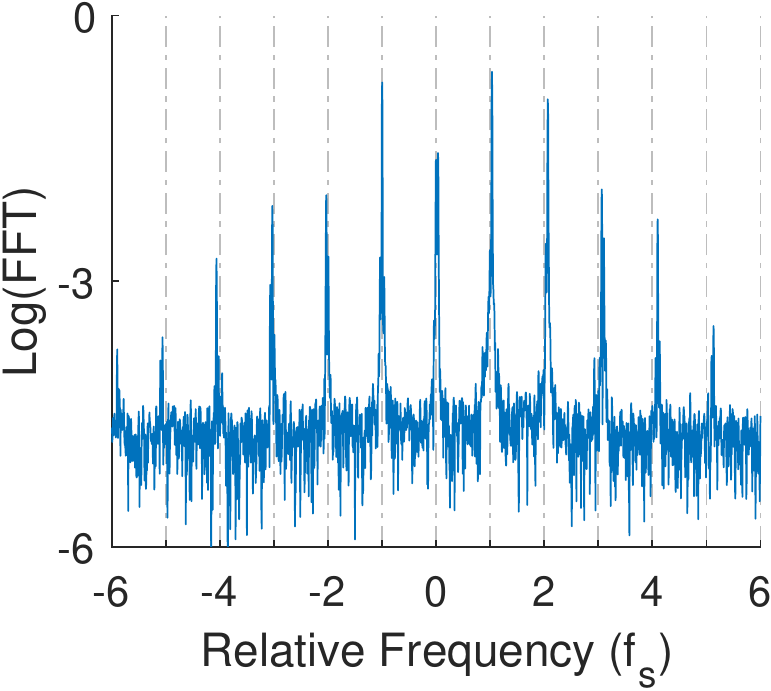}}
\caption{\label{fig:breath_6}\textit{Dark solitary wave periodic breathing.} Reconstructed phase data demonstrating the solitary wave nature of a single dark solitary periodic breather. (a) Phase vs power for a single round trip with the characteristic sharp phase jump at the dark solitary wave peak. (b) Power spectrum about the carrier frequency and scaled by the comb spacing. A well developed four-wave mixing driven frequency combs with two equi-powered eigenmodes, a characteristic of dark solitary waves in a feedback ring, is evident.}
\end{figure}

A maximum amplitude round trip for a center and edge breath are shown in~\fref{fig:breath_5b} as dotted-blue and crossed-red curves, respectively. The solitonic nature of the notch is confirmed by the fit shown in~\fref{fig:breath_5c} the solid red line is now a fit to the generic dark soliton NLS solution, given early in~\fref{eqn:dark_soliton}. A random example of a maximum amplitude round trip is shown here, but fits were made to all 870 identified peaks. A reduced $\chi^2$ of $1.09 \pm 0.42$ with a reduced $\hat{R}^2$ of $0.98 \pm 0.01$ and Jacobi-elliptic parameter, $m$ of $0.95 \pm 0.01$ were measured for the fits to the generic dark soliton solution. A Jacobi-elliptic parameter that close to unity indicates that the peaks are nearly identical to the ideal hyberbolic solution, with only minor edge effects. We note the presence of a small secondary notch located at the edge (center) of the ring for center (edge) breaths. This peak is expected as there must always be a minimum of two notches within the feedback ring to maintain phase continuity without an additional background linear phase increase of $\pi$, as discussed previously in~\sref{sec:eandm}.

The dark solitary wave nature of the underlying signal was further confirmed by the spectral and reconstructed phase data collected at 40~Gsamples/s that is shown in~\fref{fig:breath_6}. The data presented here was gathered at the same external field as the data shown in~\fref{fig:breath_5}, but at a slightly higher ring gain. This was to enhance the amplitude of the secondary peak and illustrate the ring phase continuity across multiple dark solitary wave phase shifts.~\Fref{fig:breath_6a} shows the reconstructed phase as red dots and the carrier wave envelope as solid blue. Here we see the characteristic near~$\pi$ jump for a dark solitary wave with a zero minimum, and lower jumps for smaller notches. The small (order 10 degree) difference in phase across the entire round trip is accounted for by the time delay in the electronic feedback loop, as discussed in~\sref{sec:eandm}. Also note that the phase across the peaks is not linear, indicating that the alternative description of several bright solitary waves propagating around the ring is not sufficient. The spectral data given in~\fref{fig:breath_6b} likewise shows the characteristic feature of dark solitary waves propagating within AFSs: two principal eigenmodes of equal amplitude generating a frequency comb via four-wave mixing~\cite{scott2005}.

The formation of dark solitons in an AFR geometry which supports attractive nonlinearity has previously been shown to be possible via sufficient nonlinear losses~\cite{scott2005}. The field studied here, 446~Oe, does not forbid the three-wave splitting process, as discussed early in~\sref{sec:eandm}. And as in the previous case for bright solitary periodic breathers (see~\sref{sec:bre:clean:bright}), three-wave mixing can be considered an additional source of nonlinear loss. We also remind the reader that these experiments were designed to maximize the time waves spend propagating within the ring by maximizing transducer separation. This design choice explicitly requires higher ring excitation powers to compensate for the major linear loss mechanisms present in the ring. This maximizes the influence of any nonlinear effects, including damping, on the evolution of the wave. As we approach the ring power which forbids three-wave mixing, from below, we expect to maximize the effect of that additional nonlinear-loss mechanism on the solitary wave evolution.

These considerations, taken with the previous observation that peridic breathing occurs at AFR powers below that which support full spectral comb development, account for dark solitary wave periodic breathers at field strengths above those of bright solitary wave breathers.

\section{\label{sec:bre:multi}Multi-periodic Breathing}

Multi-periodic breathing is defined by bright or dark solitary waves which breath at two or more commensurate frequencies. As with periodic breathers, discussed throughout~\sref{sec:bre}, the breathing behavior is characterized by the oscillation of scaled SWE solitary wave amplitude between zero and one accompanied by a predictable and smooth relocation within the active feedback ring. In this case it will only be necessary that these two events occur once over the longest breathing period, as opposed to during each breath. All data presented in this section was recorded on the same YIG thin film, and on the same day, as the periodic breathers discussed in sections~\ref{sec:bre:clean:dark} and~\ref{sec:bre:clean:bright}. Transducer separation was likewise the same, fixed at $1.233 \pm 0.040~\mathrm{cm}$. Envelope data was collected at $4~\mathrm{Gsamples/s}$ through a diode with a quadratic voltage response.

\begin{figure}
\subfloat[][\label{fig:breath_7a}]{\includegraphics[width=1\columnwidth]{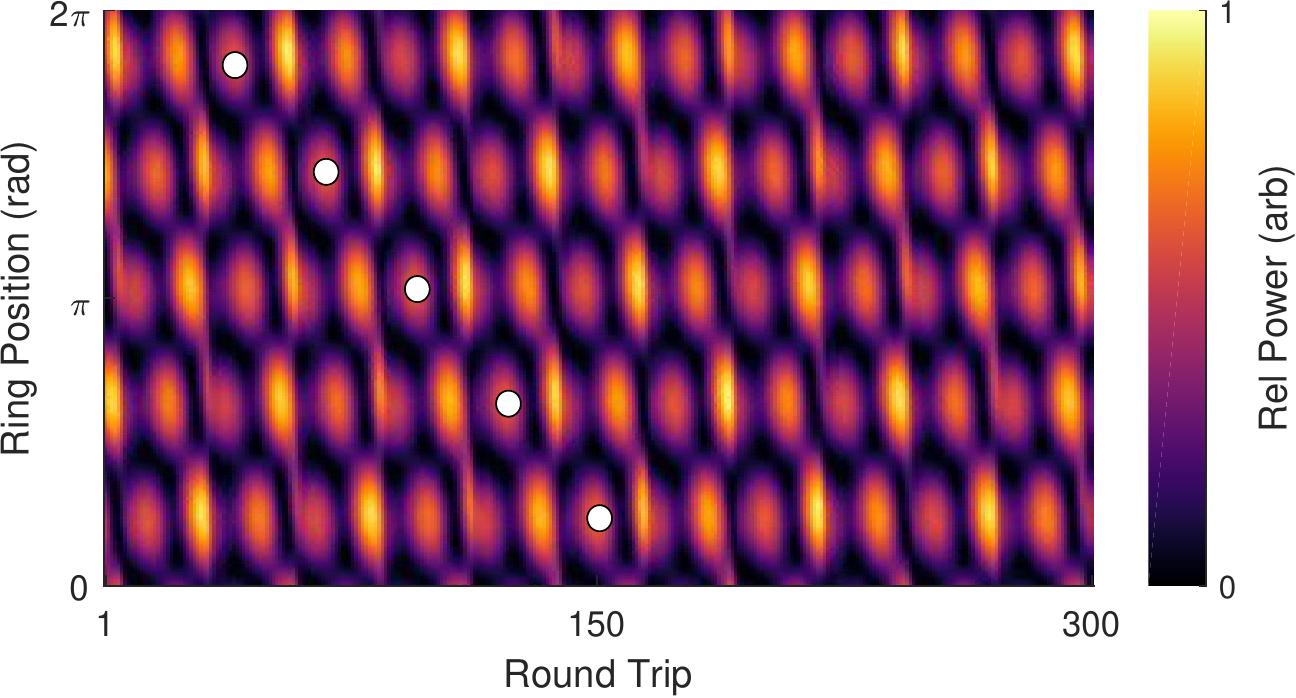}}\\
\subfloat[][\label{fig:breath_7c}]{\includegraphics[width=.49\columnwidth]{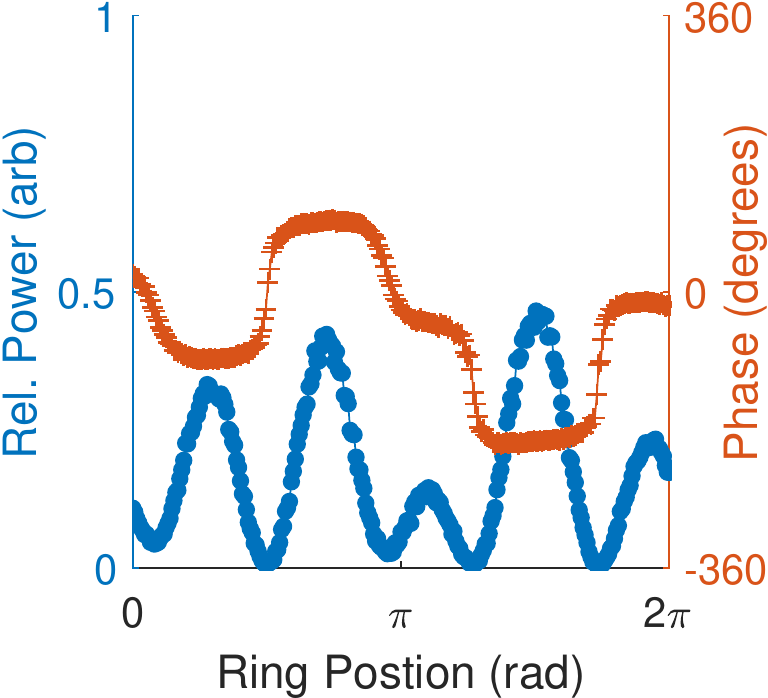}}~
\subfloat[][\label{fig:breath_7d}]{\includegraphics[width=.49\columnwidth]{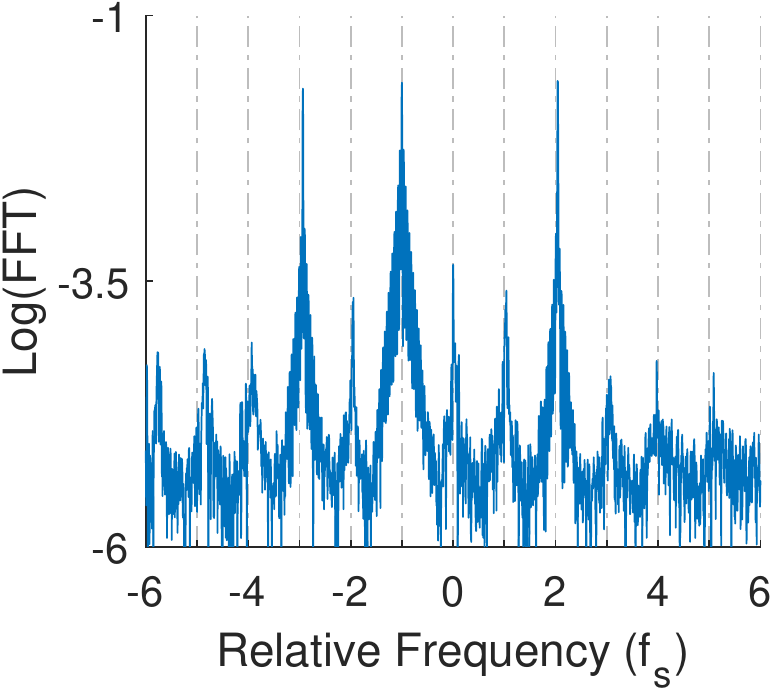}}
\caption{\label{fig:breath_7}\textit{Bright Solitary Wave multi-periodic breathing.}  (a) Reconstructed spatiotemporal plot of multi-periodic breathing for five bright solitary waves. Five periods of the principle breathing are shown with white circles. (b) shows the amplitude and phase of a single round trip in dotted blue and crossed red, respectively.  Flat phase across the wave peaks is an indication of their solitary wave nature. (c) Power spectrum relative to the carrier frequency and scaled by the comb frequency spacing. The three main peaks peaks associated with eigenmodes which have broadened due to three-wave mixing. The development of a frequency comb from four-wave mixing is also evident, and the fixed frequency spacing of these peaks is shown by the dotted grey lines.}
\end{figure}

An example of five bright solitary waves breathing at four frequencies is shown in~\fref{fig:breath_7}. A total of 8.2~$\mathrm{ms}$ of data was collected at a saturating external field strength of 383~$\mathrm{Oe}$. A round trip of 1372.82 samples was identified graphically, corresponding to a round trip time of $342.2~\mathrm{ns}$. A total of $23,892$ round trips were recorded. Group velocity amplitude may be easily estimated using the round trip time and fixed transducer separation, yielding $-3.59\times10^{-3}\pm 1.02\times10^{-4}~\mathrm{cm/ns}$. Finally, coefficients for dispersion and nonlinearity were determined by fitting to observed dispersion curves, giving values of $\mathrm{D}=1.73\times10^{-6}~\mathrm{cm^2rad/ns}$ and $\mathrm{N}=-7.23~\mathrm{rad/ns}$, respectively.

Four distinct frequencies of breathing with a common period of 238.7 round trips were identified via the spectral analysis of the envelope data. The associated breathing periods are found to be 17.04, 26.52, 47.69 and 59.98 round trip times, meaning a total of 14, 9, 5 and 4 cycles are required, respectively, to complete a common breath of 238.7 round trips. White circles in the reconstructed spatiotemporal chart, shown in~\fref{fig:breath_7a}, illustrate five iterations of the highest power breathing frequency, given by a $2\pi/5$ relocation of the solitary wave train every 26.52 round trip times. The fastest breathing period, occuring every 17.04 round trips, corresponds to the peak modulation time at any fixed ring position. The remaining two periods correspond to the variability in peak profiles across round trips at a fixed ring position and are difficult to isolate visually. It is worth highlighting the non commensurate nature of the breathing frequencies. While each breathing behavior will complete the number of cycles listed above over the duration of single common breath, that does not suggest a total recurrence of the initial state. The common breath is the minimum number of round trips required to complete an integer number of each of the four breathing periods. For example, the $2\pi/5$ relocation event completes a trip around the ring every 5 of its periods, but will complete 9 such relocations during a common breath resulting in a $2\pi/5$ shift relative to the initial condition.

A single round trip of the bright multi-periodic breather is shown in~\fref{fig:breath_7c} where crossed-red is phase in degrees, and dotted-blue is scaled amplitude. Binned data is marked and solid curves are provided as a guide for the eye. Five distinct bright solitary waves may be identified as present within the AFR. Flat phase across the bright SWE strongly suggests that the waves are solitonic in nature.

Individual peaks were not isolated. However, individual round trips were best fit by sums of the general bright solitary wave solution to the NLS with Jacobi-elliptic parameters, $m$, ranging from 0.20 to 0.9. These fits had reduced $\chi^2$ on the order of 10, but were significantly better than fits of sums of hyberbolic and sinusoidal functions where reduced $\chi^2$ was routinely above 50. All three options have mean reduced $r^2$ above 0.95.  Each individual peak maintains its qualitative spatial and temporal profiles while breathing.

While the external field strength studied here, $383~\mathrm{Oe}$, yields a spin wave power which is is too low to forbid the three-wave mixing process, we can be certain that the dynamics present within the AFR are dominated by the four-wave mixing process. The frequency spectrum relative to the carrier frequency, $2.644~\mathrm{GHz}$, reconstructed via data collected at $40~\mathrm{Gsamples/s}$, is given in~\ref{fig:breath_7d}. Here the frequency axis has been scaled by a characteristic frequency comb spacing, which is generated via the conservative four-wave mixing of ring eigenmodes, $f_\mathrm{s}=2.92\times10^{-3}~\mathrm{GHz}$. The dashed grey lines indicate the position of integer multiples of $f_\mathrm{comb}$ from the carrier frequency, and match the locations of the spectral peaks. We highlight, as discussed in \sref{sec:eandm}, that the spectral distance between ring eigenmodes is not constant, so the presence of an equi-spaced frequency comb is characteristic of four-wave mixing dominating the dynamics. The frequency comb present in this case has much broader peaks than that of a single period bright soliton breather, see~\fref{fig:breath_2b}, which is indicative of the presence of a third ring eigenmode (higher ring power) and in general with increased dynamical complexity. This type of peak broadening is most likely attributable to three-wave processes acting at higher ring powers, as this behavior has 220\% increased average round trip rms power than the single frequency case. Also note that this comb spacing corresponds to a round trip time of $341.67~\mathrm{ns}$ which is within 1\% of our previously estimated value.

\begin{figure}
\subfloat[][\label{fig:breath_7b}]{\includegraphics[width=1\columnwidth]{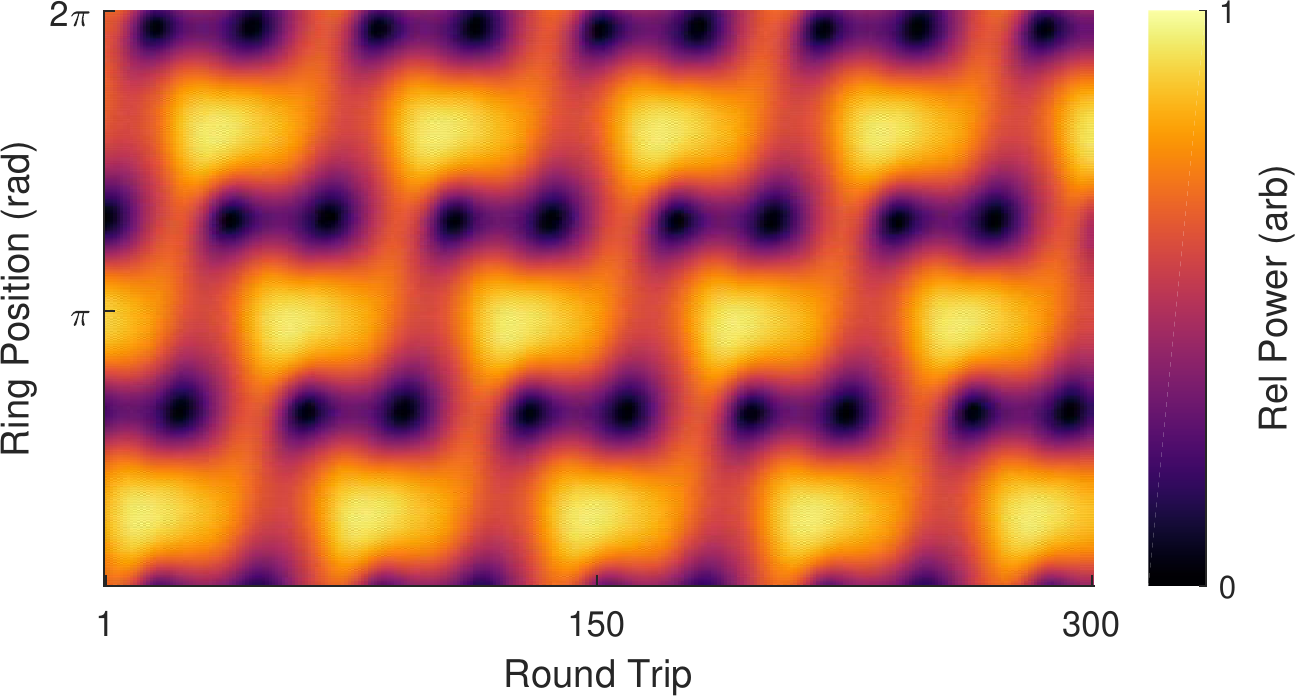}}\\
\subfloat[][\label{fig:breath_7e}]{\includegraphics[width=.49\columnwidth]{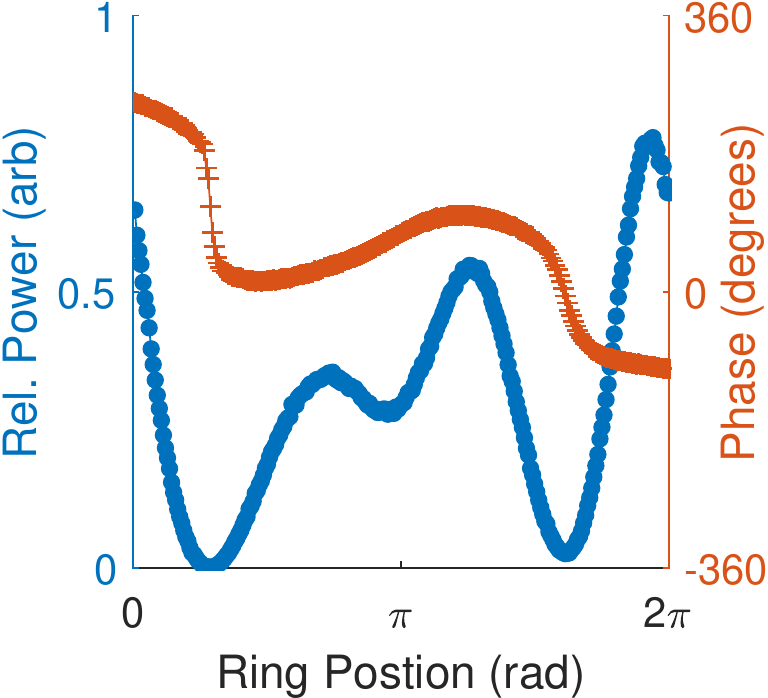}}~
\subfloat[][\label{fig:breath_7f}]{\includegraphics[width=.49\columnwidth]{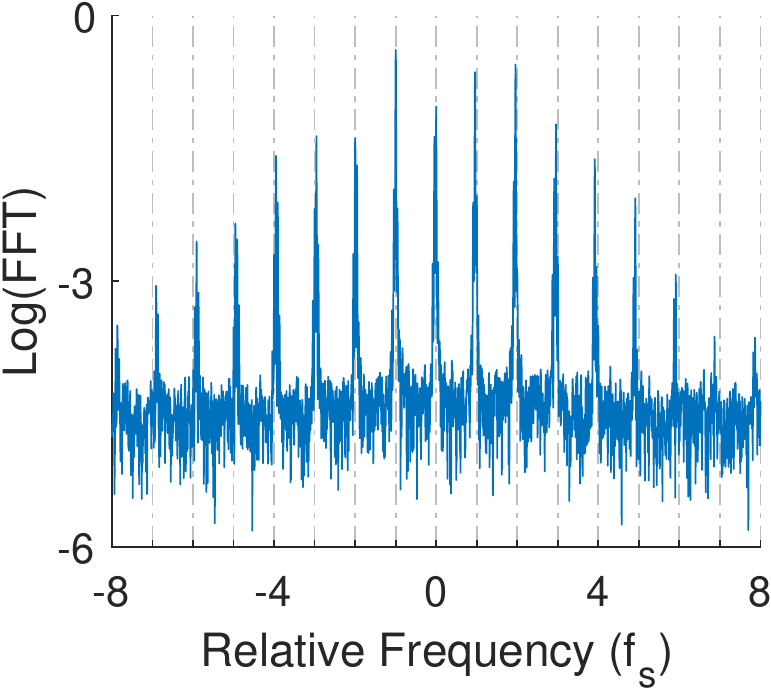}}
\caption{\label{fig:breath_8}\textit{Multi-periodic breathing of dark solitary waves.} Reconstructed spatiotemporal plot of multi-periodic breathing for three dark solitary waves.  (b) Amplitude and phase of a single round trip in dotted blue and crossed red, respectively.  Data is binned for clarity.  Near $\pi$ phase jumps across the the wave peaks is an indication of their solitary wave nature. (c) Power spectrum showing two peaks associated with eigenmodes which have developed a frequency comb from four-wave mixing. The fixed frequency spacing of these peaks is shown by the dotted grey lines.  Data shown here is about the carrier frequency and scaled by the comb frequency spacing.}
\end{figure}

Multi-periodic breathing was also observed in dark solitary waves on the same film and on the same day as the rest of the data in~\sref{sec:bre}. These dark soliton wave multi-periodic breathers were recorded with an external field strength of $460~\mathrm{Oe}$ with an observed carrier frequency of $2.941~\mathrm{GHz}$.   The round trip time was graphically identified to be 1316.484 samples by minimizing variance in group velocity across all 24,915 observed round trips. These values correspond to a round trip time of $329.12~\mathrm{ns}$ and an estimated group velocity of $-3.74\times10^{-3}\pm 1.03\times10^{-4}~\mathrm{cm/ns}$.  Dispersion and nonlinearity coefficients were also estimated via fitting to observed dispersion curves, giving values of of $\mathrm{D}=1.76\times10^{-6}~\mathrm{cm^2rad/ns}$ and $\mathrm{N}=-7.31~\mathrm{rad/ns}$, respectively. We highlight, as discussed in~\sref{sec:bre:clean:dark}, that this combination of attractive nonlinearity with dispersion does not predict the generation of dark solitary waves without the presence of higher order effects.  The carrier frequency and external field strength here likewise does not forbid the three-wave processes which are anticipated to contribute higher order losses to the peak evolution.

A reconstructed spatiotemporal plot is given in~\fref{fig:breath_7b} where three dark solitary waves can been seen undergoing multi-periodic breathing involving a $2\pi/3$ relocation as the waves evolve through time. The dark solitary wave nature of the signal can be confirmed by examining the relative phase of a single round trip, reconstructed via sampling at $40~\mathrm{Gsamples/s}$ without a diode, plotted in~\ref{fig:breath_7e}. Dotted-blue is the wave amplitude and crossed-red is phase.  A signature characteristic of dark solitary waves is seen in the near $\pi$ phase jump at minimums.

An equi-spaced spectral comb located around the carrier frequency is shown in the power spectrum plotted in~\fref{fig:breath_7f}. Here the frequency has been scaled by the characteristic comb width given by $f_\mathrm{s}=3.08\times10^{-3}~\mathrm{GHz}$, and integer multipliers of $f_\mathrm{s}$ relative to the carrier frequency are indicated with dotted-grey lines. A more robust comb structure is evident in this example than was for the single period dark solitary wave breather (see~\fref{fig:breath_6b}), driven by a 125\% increase in average round trip rms power. We note that four-wave mixing remains the principal driver of dynamical behavior with all additional spectral peaks falling on integer multipliers of $f_\mathrm{s}$.

Three noncommensurate breathing frequencies were identified via spectral analysis of the envelope. The associated breathing periods are given by 13.46, 16.82, 22.42 round trips. The minimum number of round trips necessary for all three periods to complete an integer number of periods is 67.3 and requires 5, 4 and 3 completed cycles, respectively, from the identified breathing periods. This common period corresponds to three $2\pi/3$ relocations of the maximum amplitude dark solitary wave. The remaining shorter periods correspond to shifts in the relative depth of individual dark solitary waves prior to the relocation events.

\section{\label{sec:rec}Complex Recurrence}
Recurrence is a common phenomenon in nonlinear systems and is defined by a long-period pattern or replication of a key feature. The period for recurrence is longer than any characteristic period in the problem, which distinguishes the behavior from the breathing dynamics described in sections~\ref{sec:bre}-\ref{sec:bre:multi}. The Fourier spectrum of a recurrence solution is broadband around the recurrence frequency, whereas breathers are clean. This kind of recurrence is not due to two or a few noncommensurate frequencies, but involves many conspiring factors in pattern formation -- this is why we call it complex. For example, consider two solitary waves circulating the ring with different velocities so that they overtake each other at a fixed frequency.  This is not considered complex recurrence, nor are the multi-periodic breathing dynamics of~\sref{sec:bre:multi}.  Here we look at two examples of recurrence observed in the system, one of bright solitary waves and one of dark solitary waves.  We observe varying degrees of complexity in the recurrence pattern, as well as frequency.

We note explicitly that our working definition for recurrence here excludes other types of recurrence that have been previously reported within AFR such as Fermi-Pasta-Ulam recurrence where two clean solitary waves undergo periodic recombination. This is a choice we make consciously, as we are describing a family of increasingly complicated dynamical behaviors that begin at a complexity stage above that of co-propagating solitary waves. Moreover, these dynamics appear to occur at ring gains, fields, and powers which do not support stable bright solitons.  Rather they occur at a crossover point at lower ring powers where novel dynamics are enabled by the peak broadening effects of three-wave mixing but dynamics are still principally driven by four-wave mixing.

\begin{figure}
\subfloat[][\label{fig:breath_8a}]{\includegraphics[width=1\columnwidth]{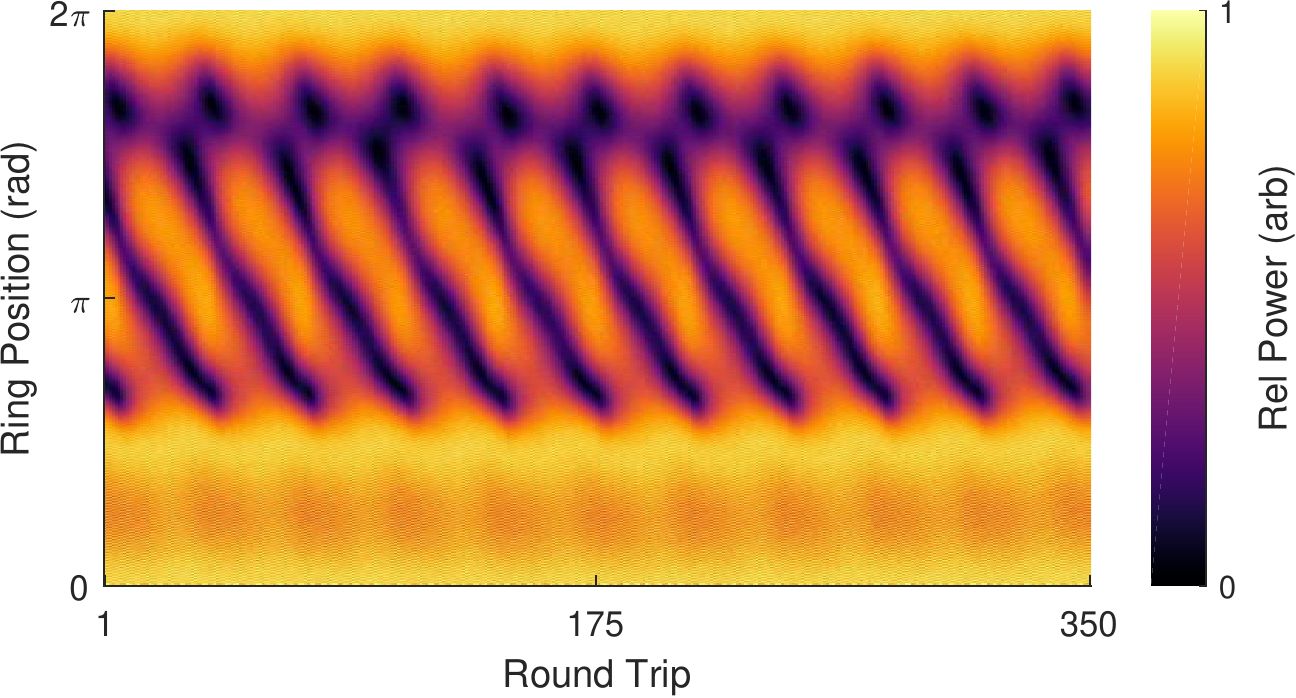}}\\
\subfloat[][\label{fig:breath_8b}]{\includegraphics[width=1\columnwidth]{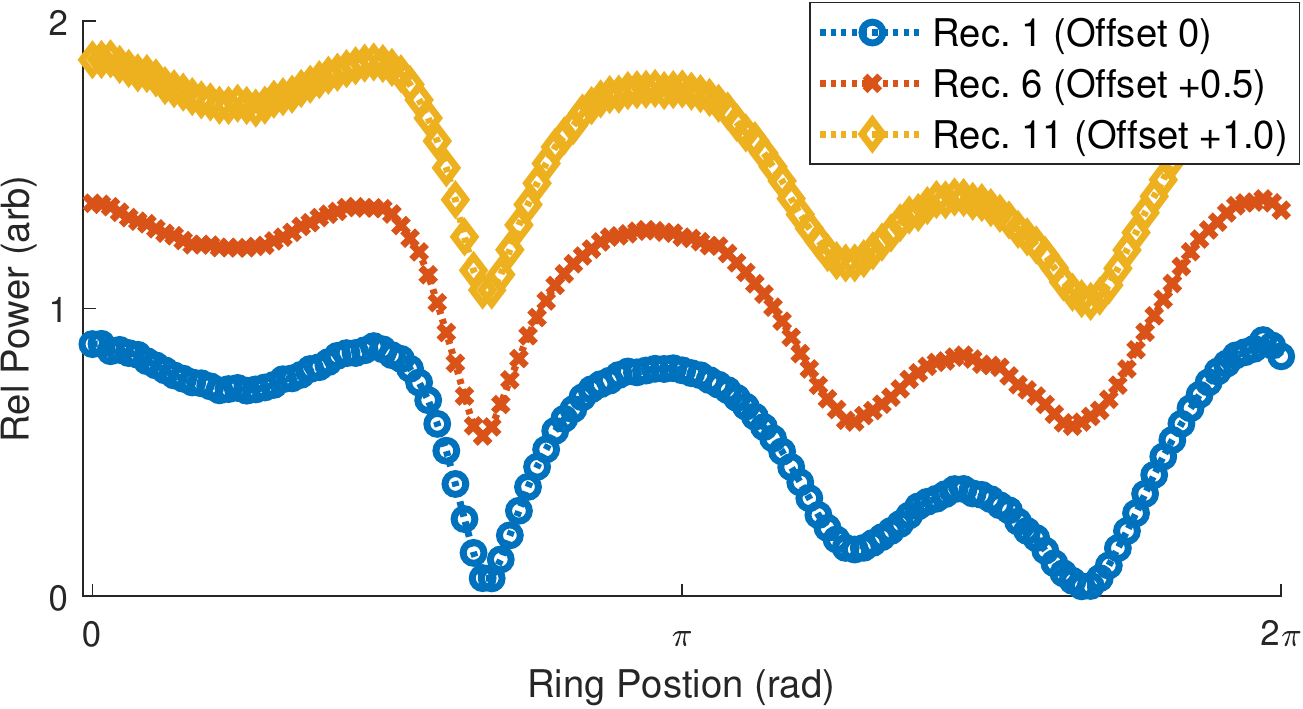}}
\caption{\label{fig:breath_9}\textit{Dark solitary wave recurrence.} (a) A reconstructed spatiotemporal plot of three dark solitary waves undergoing a complex recurrence. The nontrivial repeating shape can be identified easily, as can its periodicity. (b) The single round trips of recurrence 1, 11 and 16. These plots have been shifted by 0, 0.5 and 1 in power, respectively, to aid with direct comparison. Actual relative errors between these events is on the order of 10\% which is expected since our resolution is by the round trip, and we may not capture the exact recurrence trip.}
\end{figure}

The recurrence of a dark solitary wave is shown in~\fref{fig:breath_8a} in a reconstructed spatiotemporal plot. This particular data set has a recurrence time of 34 round trips. The first, sixth, and eleventh recurrence event are shown in~\fref{fig:breath_8b} with a vertical offset for later recurrence events to allow for a direct visual comparison.  The relative error between these recurrence round trips is on the order of 5\%, which is not unexpected since our resolution is limited to a round trip and its unlikely we'll fully observe the entire recurrence event.

A round trip time of 1317.374 samples was identified graphically, or $329.34~\mathrm{ns}$, slightly slower than that reported in the multi-periodic solution --- but well within our margins of error. 8.2ms of data was collected, or 24897 round trips and 730 distinct recurrence events.

This behavior was observed under the same conditions as the multi-periodic dark solitary wave breather, above, only at a higher total ring gain resulting in an average RMS round trip power 13\% greater than in the previous example.  The field strength and carrier frequency remain the same.

A shift from multi-periodic breathing into a more complex dynamic such as recurrence as ring gain is increased is not surprising. Higher ring gains result in additional eigenmodes circulating the thin film, and further, since this data was collected in a regime where three-wave mixing was not forbidden, we do expect higher powers to result in spectral peak broadening.  This example was included here not because of its particularly complex nature, but rather as an illustrative example of how ring dynamics increase in complexity as the spin wave power increases.  The progression from a single solitary wave periodic breather, to multi-periodic multi-peak breathers ultimately into the broadband reccurence shown here is natural.

\begin{figure}
\subfloat[][\label{fig:breath_9a}]{\includegraphics[width=1\columnwidth]{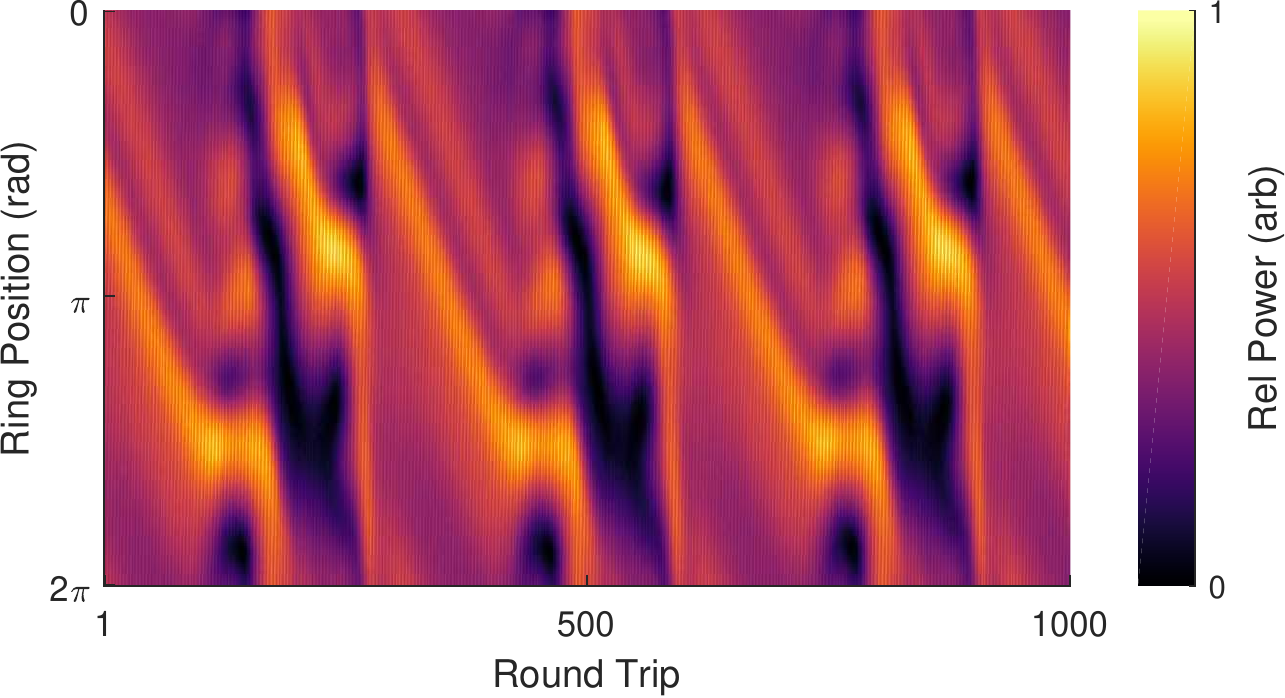}}\\
\subfloat[][\label{fig:breath_9b}]{\includegraphics[width=1\columnwidth]{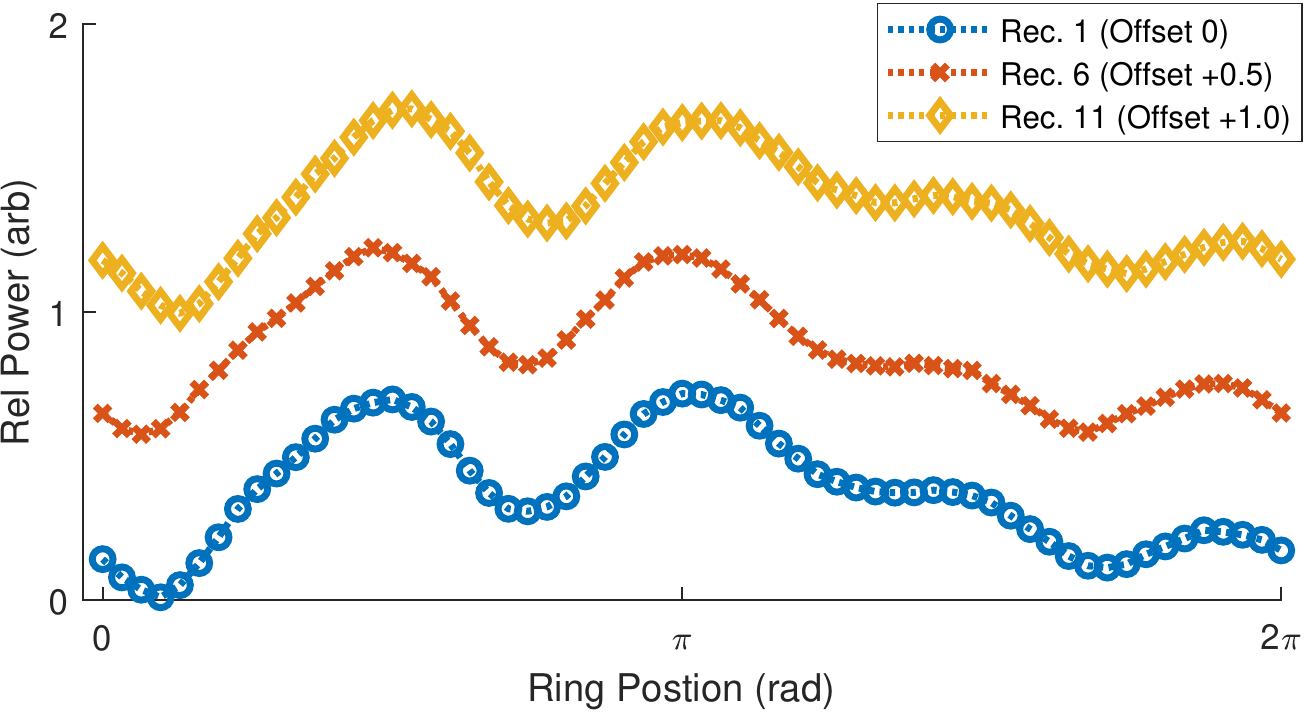}}
\caption{\label{fig:breath_10}\textit{Bright solitary wave recurrence.} (a) Four bright solitary waves undergoing a complex recurrence. Format is the same as \fref{fig:breath_9}.  The nontrivial repeating shape can be identified easily, as can its periodicity. (b) The single round trips of recurrence 1, 11 and 16. These plots have been shifted by 0, 0.5 and 1 in power, respectively, to aid with direct comparison. }
\end{figure}

At higher ring gains and fields even more complex recurrence events are common. An example of this with five bright solitary waves circulating the ring is shown in~\ref{fig:breath_9a} where a reconstructed spatiotemporal plot shows three periods of a highly intricate recurrence event. This behavior exhibits a recurrence every 316.6 round trips. The individual round trips for recurrences 1, 6 and 11 are plotted in~\fref{fig:breath_9b} with power offsets to allow for the visual comparison. The percent error between the first recursion and the 6th and 11th recursions is on the order of 5\% at each point. This is not unexpected since our resolution is limited to individual round trips and the actual recursion time is not an integer.

This data was collected at a field strength of $773~\mathrm{Oe}$ and a sampling rate of $1~\mathrm{Gsample/s}$ through a diode with a quadratic response to voltage.  A round trip time of $245.485~\mathrm{ns}$ was identified graphically by minimizing the variation of the group velocity over the recorded 66,806 round trips covering $16.4~\mathrm{ms}$. The transducer separation was measured to be $1.147~\mathrm{cm}$ with a photo microscope. An estimate of group velocity may obtained from the round trip time and the transducer separation, giving $-4.68\times10^{-3}~\mathrm{cm/ns}$ which closely matches the group velocity obtained from recorded dispersion data of $4.77\pm0.32\times10^{-3}~\mathrm{cm/ns}$.  Dispersion and nonlinearity coefficients were also obtained via fitting to recorded dispersion data, giving $D=2.9\times10^{-6}\pm 1.8\times10^{-7} ~\mathrm{cm^2rad/ns}$ and $N=-8.72~\mathrm{rad/ns}$ both in close agreement with other experiments of this kind.

\section{\label{sec:sss}Spontaneous Spatial Shifts}
The most dramatic dynamical behavior observed is the spontaneous spatial shift. This behavior is characterized by quick, unpredictable movement of an otherwise underlying behavior from one place in the ring to another (in the group velocity frame). Unique from the previous two examples of periodic breathing and complex recurrence, spatial shifting does not occur at a fixed frequency and involves an abrupt, unpredictable transition rather than smooth, periodic relocation or recombination.

\begin{figure}
\includegraphics[width=1\columnwidth]{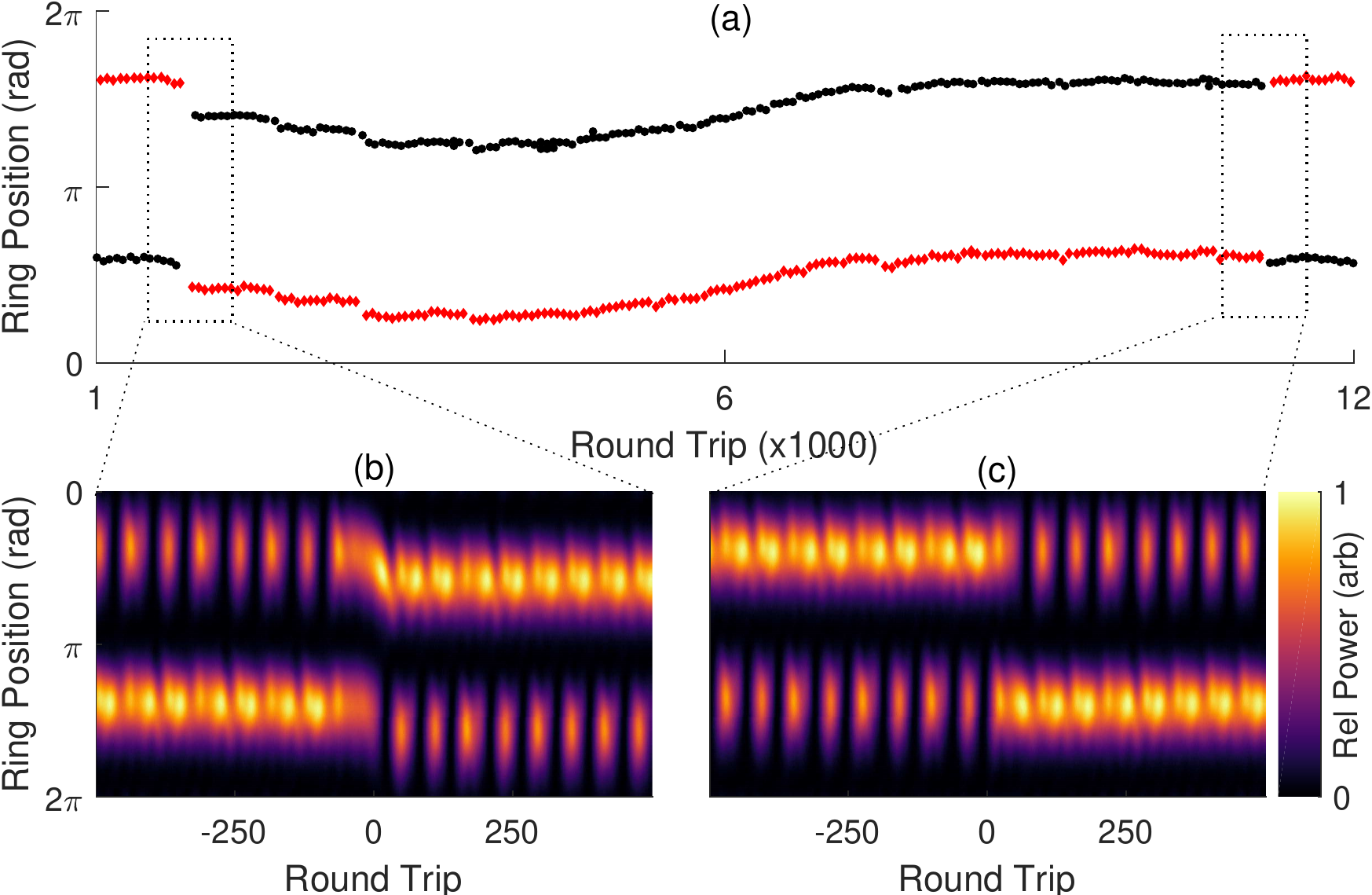}
\caption{\label{fig:sss_1}\textit{Spontaneous Spatial Shifts} (a) All peaks detected for a spontaneous spatial shifting behavior. Red dots indicate the short peaks which oscillate between a normalized magnitude of 0.7 and 0, while black dots show the high peaks which oscillate between 0.7 and 1. (b) Two shifting events when the relative location of the two stable modulating solitons abruptly shifts within the ring. 
}
\end{figure}

An example of spontaneous spatial shifts is shown in~\fref{fig:sss_1} where two spatial shifting events are observed over 12,000 round trips. The otherwise stable underlying dynamic which undergoes the spontaneous spatial shifting is two co-propagating bright solitary waves undergoing periodic modulation. A high amplitude main peak modulates alongside a smaller side peak. The periods of the modulation are the same for each peak, 61.19 round trip times, but are out of phase. The smaller peak modulates from a scaled power of 0 to 0.7, while the larger peak modulates from a scaled power of 0.7 to 1. Note that the smaller peak is not undergoing the periodic breathing dynamic discussed earlier in~\sref{sec:bre}, as it does not undergo any predictable relocation within the ring. This behavior was also predicted by our previous numerical phase space exploration~\cite{anderson2014complex} where we refer to it as ``complex co-propagation.'' We found it occurred when all losses present within the AFR were roughly of the same order. This means we expect this dynamical behavior to exist when additional losses from three-wave splitting are forbidden, as is the case here where the external field strength was 565~$\mathrm{Oe}$ resulting in a operating frequency of 3.37~$\mathrm{GHz}$. This is above the minimum frequency required to forbid three-wave splitting as discussed in~\ref{sec:eandm:SW}. \Fref{fig:sss_1}a shows the isolated peak locations of the high amplitude modulating solitary wave as black dots and of the low amplitude solitary wave as red diamonds. The two spontaneous spatial shifts are isolated into reconstructed spatiotemporal plots b and c. Both relocation events occur over less than 100 total round trips, with the first event being the less clean of the two.  The stability of the underlying complex co-propagation before and after the spontaneous spatial shifts is evident.

This data were collected at a sampling rate of $4~\mathrm{Gsample/s}$ and a round trip time of $324.917~\mathrm{ns}$ was identified graphically by minimizing the variation of the group velocity over the recorded 25,236 round trips covering $8.2~\mathrm{ms}$. The transducer separation was measured to be $1.232~\mathrm{cm}$. An estimate of group velocity can be obtained from the round trip time and the transducer separation, giving $-3.9\times10^{-3}~\mathrm{cm/ns}$ which closely matches the group velocity obtained from recorded dispersion data of $-3.79\pm0.21\times10^{-3}~\mathrm{cm/ns}$.  Dispersion and nonlinearity were $D=1.9\times10^{-6}~\mathrm{cm^2rad/ns}$ and $N=-7.85~\mathrm{rad/ns}$. The peaks for both the main and side peaks were isolated and used to determine the jitter in group velocity showing of change of $0.96 \pm 2.6 \times 10^-4 \mathrm{cm/ns}$ per round trip. This amounts to a less than 1\% change in group velocity per round trip, and the jitter is roughly normally distributed with an Anderson-Darling statistic of 2.5 where normality is formally rejected at a level of 0.75. 402 major and minor peaks at maximum amplitude were best fit to a Jacobi-elliptic cn with a $\chi^2=0.29\pm1$ with a parameter $m=0.97\pm0.11$ and a mean adjusted $\mathrm{r}^2=.99\pm0.01$ confirming the bright solitary wave nature of the underlying dynamic.

\section{\label{sec:int}Intermittency}
Intermittency is common in complex dynamical systems and occurs when two or more nearby attractors in phase space overlap. This results in the seemingly random oscillation between two or more distinct dynamical behaviors as the systems moves between attractors.  It is important to note that these attractors need not ``physically'' overlap in phase space, but that drift in experimental parameters such as heating or noise can result in shifts of the nearby basins of attraction that result in dynamical changes.

\begin{figure}[!ht]
\subfloat[][\label{fig:int_1a}]{\includegraphics[width=1\columnwidth]{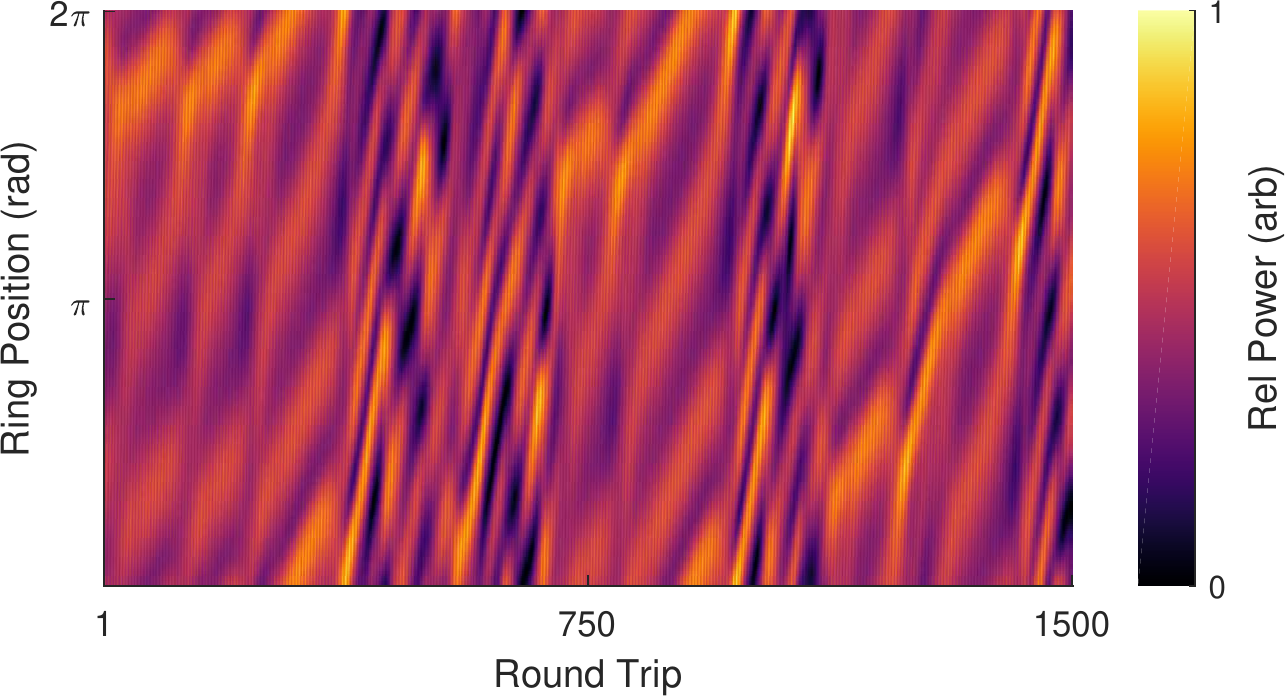}}\\
\subfloat[][\label{fig:int_1b}]{\includegraphics[width=.49\columnwidth]{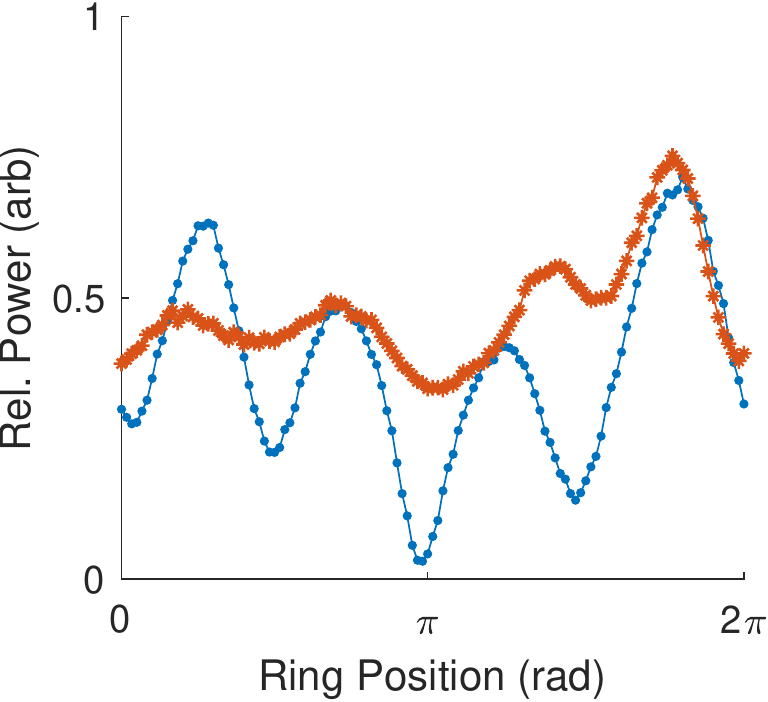}}~
\subfloat[][\label{fig:int_1c}]{\includegraphics[width=.49\columnwidth]{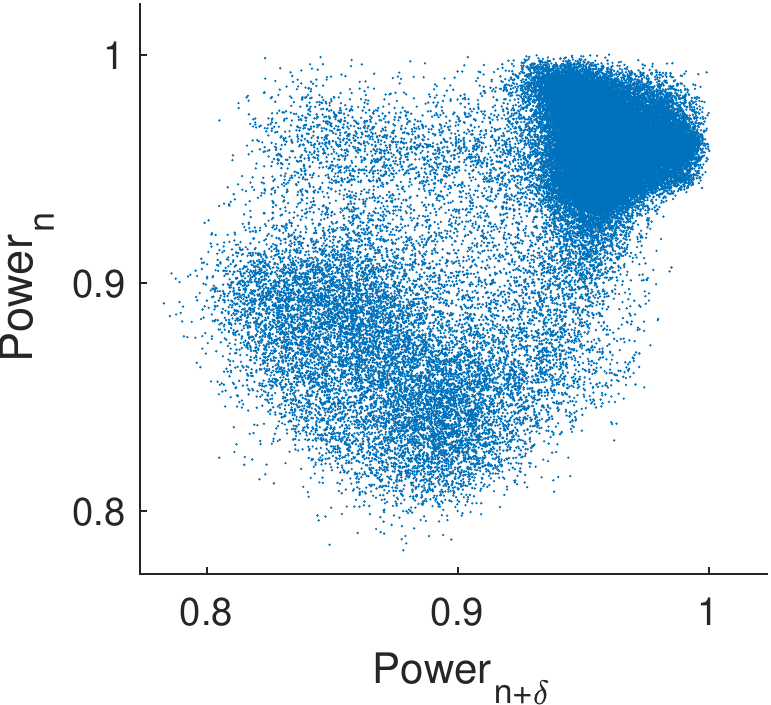}}\\
\subfloat[][\label{fig:int_1d}]{\includegraphics[width=1\columnwidth]{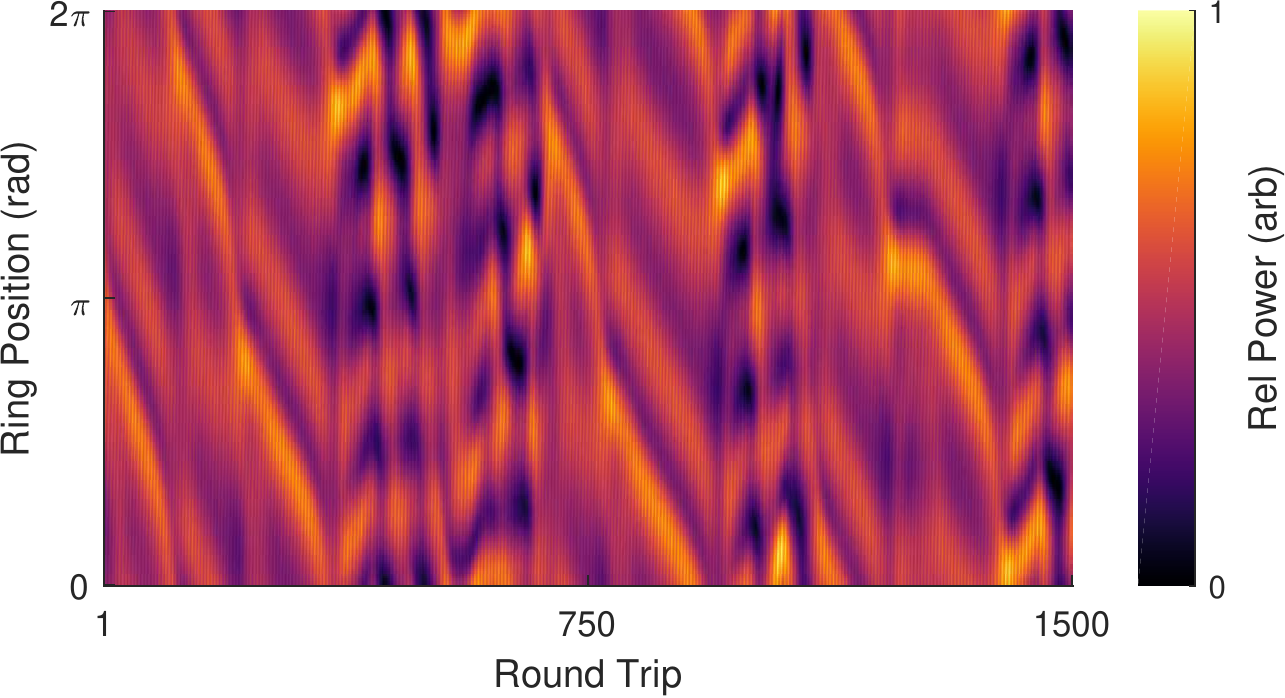}}
\caption{\label{fig:int}\textit{Intermittency.} (a) Reconstructed spatiotemporal plot showing the unpredictable shifting between two stable dynamical behaviors. The two distinct dynamics are shown in (b) with four interacting solitary waves (dotted-blue points) and 4 waves propagating on a floor of scaled power 0.4 (crossed-red points). (c) Attractor reconstruction with round trip rms power, $E(\mathrm{n})$. The delay, $\delta$, is chosen as the first minimum of the autocorrelation function and $E(\mathrm{n})$ vs $E(\mathrm{n}+\delta)$ is shown. Data from 66833 recorded round trips shows the qualitative presence of two overlapping attractors, with a majority of the time spent in the higher rms power state shown as the crossed-red behavior in plot (b). (d) Spatiotemporal plot in the group velocity frame of the \emph{second} dynamic.}
\end{figure}

A typical example of intermittency is shown in~\fref{fig:int}, where~\fref{fig:int_1a} is a reconstructed spatiotemporal plot showing 1500 round trips spanning 3 regions for each of the two underlying dynamical behaviors. The first behavior, shown in crossed-red points in~\fref{fig:int_1b}, is a principal peaks periodically modulating with 3 smaller peaks on a floor of relative power 0.4. This type of evolution was also numerically predicted to be observable in this system in previous numerical work~\cite{anderson2014complex}, where it was called ``asymmetrical interactions'' on an energy 'plateau' whose amplitude satisfied the energy balance of the CQCGL equation $L+C|u|^2+Q|u|^4=0$.  \Fref{fig:int_1a} is reconstructed in the group velocity frame of this dynamic, corresponding to a round trip time of 243.94~$\mathrm{ns}$. Of the 66,000 recorded round trips the ring evolved in the crossed-red behavior for 81\% of them. The average occupation duration for this dynamic was 1053 round trips, but varied from 200 up to 3000. It is shown propagating within the ring from round trips 1-357, 682-961 and 1096-1401 in~\fref{fig:int_1a}. \Fref{fig:int_1d} is the spatiotemporal plot of the same 1500 round trips as in~\fref{fig:int_1a} but in the group velocity frame of the second dynamic shown as the dotted-blue line in~\fref{fig:int_1b}. The second behavior has a round trip time of 245.2~$\mathrm{ns}$ and propagates the ring for an average of 260 round trips. The maximum observed continuous lifetime of this behavior was 350 round trips, the ring existed in this regime for 19\% of the recorded round trips. The second dynamic propagates within the ring from round trips 357-682, 961-1096 and 1401-1500.  This second behavior is four solitary waves interacting. No clear, stable breathing behavior is observed but the waves do modulate and move about the ring. A total of 102 transitions were observed in the 66,000 recorded round trips of data, corresponding to roughly over 50 distinct propagation periods for each of the two dynamical behaviors.

\Fref{fig:int_1c} shows an attractor reconstructed from the RMS $n$th round trip power, $E_{\mathrm{rms}}(n)$, via phase space time-delay embedding. Time-delay embedding process involves creating "independent" vectors from a single recorded time series using a delay which minimizes the correlation between success vectors defined as $[E_{\mathrm{rms}}(n), E_{\mathrm{rms}}(n+\delta) , E_{\mathrm{rms}}(n+2\delta) , ... , E_{\mathrm{rms}}(n+m\delta) ]$ where $\delta$ is the delay constant and $m$ is the maximum number of vectors which can be constructed for a finite length time series.  The appropriate choice of the delay constant is an active area of debate, but for our qualitative purposes the first zero crossing of the autocorrelation function of $E_{\mathrm{rms}}(n)$ will minimize the linear correlation between successive vectors. Additional discussion of nonlinear time series analysis is well beyond the scope of this paper, and curious readers are directed to these fine reviews as introductions to the field~\cite{Bradley2015,Kantz:2004,Abarbanel1996}.~\Fref{fig:int_1c} shows, qualitatively, the existence of two distinct attractors, where the density of points equates to their relative occupation frequency.  The most common ring behavior (crossed-red points in~\fref{fig:int_1b}) has an average scaled RMS round trip power near 1. The second behavior (dotted-blue points) has an average scaled RMS round trip power near 0.9. These numbers are in complete agreement with RMS round trip powers determined by isolating each individual behavior's time within the ring. The relative number of points in the upper right attractor, corresponding to the first behavior, is 77\%, while the fraction of points in the lower attractor is 17\%. This agrees closely with our relative frequencies above and also suggests roughly 6\% of propagation time is spent transitioning from one attractor to another.

This data were collected at the same field strength as the bright soliton complex recurrence above. That is given by $773~\mathrm{Oe}$ and a sampling rate of $1~\mathrm{Gsample/s}$. Round trip times for each behavior were identified graphically by minimizing the variation of the group velocity over the recorded 66,000 round trips totaling $16.4~\mathrm{ms}$. The transducer separation was fixed at $1.147~\mathrm{cm}$. Group velocity estimates are obtained from the round trip time and the transducer separation, giving $-4.52\times10^{-3}~\mathrm{cm/ns}$ and $-4.49\times10^{-3}~\mathrm{cm/ns}$ for the first and second behaviors listed above, respectively. These are again in close agreement with the group velocity determined from the experimentally measured dispersion curve, $4.77\pm0.32\times10^{-3}~\mathrm{cm/ns}$.  Dispersion and nonlinearity coefficients were $D=2.9\times10^{-6}\pm 1.8\times10^{-7} ~\mathrm{cm^2rad/ns}$ and $N=-8.72~\mathrm{rad/ns}$.

\section{\label{sec:con}Conclusions and Outlook}

We report the clean experimental realization of 10 complex behaviors across 4 distinct categories of dynamical pattern formation that were previously numerically predicted~\cite{anderson2014complex} to be observable for backward-volume spin waves (BVSWs) propagating within active magnetic thin film-based feedback rings (AFRs). These four regimes span (1) periodic and multi-periodic breathing, (2) complex recurrence, (3) spontaneous spatial shifting, and (4) intermittency.  We provided experimental examples of all these behaviors and performed quantitative analysis including reconstruction of the strange attractor underlying (4).  Besides these predicted behaviors, we also discovered dynamical pattern formation for dark solitary waves which evolve under attractive instead of repulsive nonlinearity. Dark solitary waves were observed in the regimes of periodic and multi-periodic breathing and complex recurrence organically via self-generation and without any external potentials, sources or other effects.

Our results confirm that the cubic-quintic complex Ginzburg-Landau equation (CQCGL) is a simple yet viable model for the study of fundamental nonlinear dynamics for driven, damped waves which propagate in nonlinear, dispersive media.  We have demonstrated that spin wave envelope (SWE) solitary waves in AFRs provide an approachable and flexible table top experiment to study the emergence of many complex regimes in nonlinear dynamics. Additionally, our experimental verification of these dynamical regimes show that such ideas are not simply theoretical but in fact occur in the real physical world and are observable in an approachable, tunable spin-wave system which matches the conditions of many other real-world physical systems and are therefore promising for technological applications. This provides for potential benchmarking and insight not only into fiber optics, hydrodynamics, and the many other fields of nonlinear dynamics with classical waves, but can provide for a classical benchmark on gain/loss open quantum systems such as attractive BEC Sagnac interferometers.

Previous studies of dark and bright solitary wave train dynamics of SWE solitary waves in AFRs explicitly occurred in regimes which forbade three-wave splitting. Our work demonstrates that four-wave mixing can remain the dominant dynamical force even when three-wave splitting is allowed for some or all of the spin wave passband. This opens up additional spin wave regimes for studying solitary wave train dynamics and additional exploration of this phase space. Similar explorations for forward volume and surface spin waves and mixed excitation should yield intriguing and important dynamical behaviors. Transient dynamics in these regimes also should be investigated.  Further exploration of additional spin wave regimes and numerical study of dark solitary wave dynamics are also warranted and verification of the existence of similar dynamical behaviors in analogous physical systems would be exceedingly valuable.

\begin{acknowledgments}
Work at CSU was supported by the U.S. National Science Foundation under Grants No. DMR-1407962 and No. EFMA-1641989.
\end{acknowledgments}

\bibliography{refs}
\bibliographystyle{unsrt}

\end{document}